\documentclass[review]{elsarticle}
\journal{Combustion and Flame}

\usepackage{hyperref}

\usepackage{amsmath}
\usepackage{amssymb}
\usepackage{caption}
\usepackage{graphicx}
\usepackage{latexsym}
\usepackage{times}
\usepackage{comment}
\usepackage{float}
\usepackage{upgreek}
\usepackage{textgreek}
\newcommand{\gmu}{\textmugreek}

\usepackage{xcolor,soul}

\begin{document}

\begin{frontmatter}
	\title{On the role of transverse detonation waves in the re-establishment of attenuated detonations in methane--oxygen}
	\author[cwru]{Grace Floring}
	\author[cwru]{Mohnish Peswani}
	\author[cwru]{Brian Maxwell\corref{cor1}}
	\ead{brian.maxwell@case.edu}
	\cortext[cor1]{Corresponding Author}
	\address[cwru]{Department of Mechanical and Aerospace Engineering, Case Western Reserve University,\\ 10900 Euclid Avenue, Glennan 418, Cleveland Ohio, 44106, USA}

\begin{abstract}

{This investigation concerns the problem of detonation attenuation in stoichiometric methane-oxygen and its re-establishment following its interaction with obstacles, using high resolution numerical simulation.  The main focus was on the role of the transverse detonation on the re-establishment of the detonation wave, and the importance of applying a numerical combustion model that responds appropriately to the thermodynamic state behind the complex shock wave dynamics.}   We applied an efficient thermochemically derived four-step global combustion model using an Euler simulation framework to investigate the critical regimes present.  While past attempts at using one- or two- step models have failed to capture transverse detonations, {for this scenario,} our simulations have demonstrated that the four-step combustion model is able to capture this feature.  We thus suggest that to correctly model detonation re-initiation in characteristically unstable mixtures, an applied combustion model should contain at least an adequate description to permit the correct ignition {and state variable} response when changes in temperature and pressure occur, i.e.~behind shocks.  Our simulations reveal that {(1) there is a relationship between the critical outcomes possible and the mixture cell size, and (2) while pockets of unburned gas may exist when a detonation re-initiates, it is not the direct rapid consumption of these pockets that gives rise to transverse detonations.  Instead, the transverse detonations are initiated through pressure amplification of reaction zones at burned/unburned gas interfaces whose combustion rates have been enhanced through Richmyer-Meshkov instabilities associated with the passing of transverse shock waves, or by spontaneous ignition of hot spots, which can form into detonations through the Zeldovich gradient mechanism. In both situations, non-uniform ignition delay times are found to play a role.  Finally, we found that the transverse detonations} are in fact Chapman-Jouguet detonations, but whose presence contributes to overdriving the re-initiated detonation along the Mach stem.

\end{abstract}
	
	\begin{keyword}
		compressible flows \sep detonation waves \sep critical re-initiation \sep four step combustion \sep numerical simulation \sep transverse detonations
	\end{keyword}

\end{frontmatter}


\section{Introduction}

In this investigation, we revisit the problem of detonation attenuation and it's re-establishment following its interaction with obstacles, which has been investigated previously both experimentally \cite{Makris1993a,Chao2006,Zhu2007,Radulescu2011,Bhattacharjee2013b,Bhattacharjee2013,Saif2016} and numerically \cite{Radulescu2011,Bhattacharjee2013,lau2013numerical,MAXWELL2018340}.  In particular, this investigation focuses on the role of the transverse detonation on the re-establishment of the detonation wave, and investigates the importance of applying a numerical combustion model that responds appropriately to the thermodynamic state behind the complex shock wave dynamics.  This problem is particularly important for the development and validation of numerical strategies to simulate and predict the final stages of deflagration to detonation transition (DDT), as critical shock-flame complex regimes may be established close to the choked Chapman-Jouguet (CJ)-deflagration velocity.  It is currently believed that a sufficient condition for DDT to occur is when flame propagation reaches this velocity limit \cite{Rakotoarison2020}.

In early experiments, transverse shock waves were believed to play an important role in the re-establishment of the detonation wave \cite{Chao2006,Zhu2007}.  In Radulescu and Maxwell \cite{Radulescu2011}, transverse detonations were observed during the re-establishment of detonation waves in acetylene--oxygen, yet such a feature could not be captured numerically at the time.  We note here that this transverse detonation is also a key component in critical regimes of detonation diffraction \cite{Edwards1979} and also in marginal or spinning detonation propagation \cite{Voitsekhovskii1969,Strehlow1974}. In more recent experiments, Bhattacharjee et al.~\cite{Bhattacharjee2013} investigated several possible mechanisms that contribute to detonation re-initiation.  In general, they found that forward jetting of combustion products behind the Mach shock was found to play an important role in triggering rapid ignition and coupling to the Mach shock.  In some critical cases, a large pocket of unburned reactive gas remained behind, while in more sensitive reactive mixtures it was believed that rapid reaction of this pocket lead to the establishment of the transverse detonation.  Although the burn-up of these pockets has not yet been formally linked to the transverse detonation feature, it has been determined through numerical simulation that the burning rate of such pockets can be influenced by the strength of transverse shock waves during detonation propagation of irregular reactive mixtures \cite{Mahmoudi2011}.  Moreover, the burning rate of these pockets has been found to influence the cellular structure in methane--oxygen \cite{maxwell_2017}.

Although Euler simulations coupled to one- or two-step Arrhenius combustion kinetics have been attempted to capture detonation re-initiation as observed in experiments \cite{Radulescu2011,Bhattacharjee2013,lau2013numerical}, recovered solutions were found to depend highly on the resolution adopted.  Moreover, self-sustained transverse detonations were never observed in these simulations.  This was, to some extent, believed to be a consequence of using simplified chemical mechanisms.  These are generally tuned to give the appropriate ignition delay for only a particular state, i.e.~ the von Neumann state, but do not correctly reproduce the detailed reaction zone structures.  Moreover, the application of the calorically perfect gas assumption also leads to significant errors in prediction of the state behind the various shock wave dynamics present.  In more recent work, which also adopted a one-step combustion modeling approach, Maxwell et al.~\cite{MAXWELL2018340} found that adequate closure of the turbulent combustion resulted in improved prediction of the re-initiation of a detonation, but also did not predict any transverse detonation features.  Recent simulations of highly irregular detonation propagation \cite{VIJAYAKUMAR2020_thesis}, however, have shown that the application of a reduced detailed elementary reaction mechanism can indeed reproduce the various re-initiation regimes as observed experimentally.  It is clear from all of this past work that in order to capture the complete set of features observed during the critical re-establishment of a detonation wave, a sufficiently detailed description of the chemical reactions is required.  Moreover, the need exists to develop low-memory and low-overhead strategies to investigate detonation phenomena at high resolution and at larger scales.

In the current study, we address the problem of simulating detonation quenching and re-initiation following its interaction with a cylindrical obstacle in methane--oxygen, as observed by Bhattacharjee et al.~\cite{Bhattacharjee2013}, by attempting to capture the transverse detonation phenomenon using a more detailed, but minimal global description of the chemistry.  Specifically, we use a thermally perfect four-step global reaction mechanism \cite{Zhu_2012}, with temperature dependent properties, which has been calibrated to reproduce methane--oxygen reaction characteristics in a wide range of temperatures and pressures \cite{Peswani_IMECE2020,Peswani_ussci2021,Peswani2022}.  Through this approach we aim to determine if such a minimal thermochemically derived combustion model can be used to capture important features of detonation initiation, i.e.~transverse detonations.  We also aim to discover the {mechanisms of formation and the roles of these waves}, and to what extent rapid burning of reactive gas pockets contribute to the formation of these transverse detonation waves.

\section{Numerical modeling approach}

\subsection{Governing equations and combustion model}

In the current study, the two-dimensional reactive Euler equations were solved, which thus explicitly ignores diffusion effects.  Instead, deflagrative burning on reaction surfaces was driven through numerical diffusion associated with the finite-volume scheme adopted.  The complete set of conservation laws for mass, momentum, total energy, and $i$th chemical species solved here is
\begin{gather}
   \frac{\partial {\rho}}{\partial {t}} + \nabla \cdot ({\rho}{\boldsymbol{u}}) = 0
   \label{eqn.mass_euler}\\
   \frac{\partial ({\rho} \boldsymbol{{{u}}})}{\partial {t}} + \nabla \cdot ({\rho} \boldsymbol{{{u}}} \otimes \boldsymbol{{{u}}}) + \nabla {p}  = 0
   \label{eqn.momentum_euler}\\
   \frac{\partial ({\rho} {E})}{\partial {t}} + \nabla \cdot \biggl(({\rho}{E}+{p})\boldsymbol{{{u}}}\biggr) = 0
   \label{eqn.energy_euler}\\
   \frac{\partial ({\rho} {Y_i})}{\partial {t}} + \nabla \cdot ({\rho} \boldsymbol{{{u}}} {Y_i}) = {\dot{{\omega_i}}},
   \label{eqn.reactant_euler}
\end{gather}
where $\rho$, $\boldsymbol{u}$, $p$, $Y_i$, ${\dot{{\omega_i}}}$, refer to the density, velocity vector, pressure, mass fraction of the $i$th species, and the reaction rate of of the $i$th species, respectively.  The total specific energy for a thermally perfect gas is given by
\begin{equation}
    E= \sum{(Y_i h_i)}-\frac{p}{\rho} + \frac{1}{2}|\boldsymbol{u}^2|,
\end{equation}
where $h_i$ is the enthalpy of the $i$th species, and the temperature ($T$) is determined by the ideal gas law,
\begin{equation}
    p=\rho RT,
\end{equation}
where $R$ is the specific gas constant.  Finally, the speed of sound is computed using the chemically frozen ratio of specific heat capacities, $\gamma=c_p/c_v$, through
\begin{equation}
    c^2=\gamma {p}/{\rho}.
\end{equation}
The specific heat capacities,  $c_p$ and $c_v$, and enthalpies for each species, $h_i$, are determined by the usual temperature dependent NASA polynomial approximations \cite{kee1996chemkin} for a multi-component gas. Since {complete} detailed hydrocarbon chemistry descriptions are not amenable to high resolution simulations, we instead applied a thermochemically derived four-step global reaction mechanism \cite{Zhu_2012}, which has been calibrated to reproduce various constant-volume and one-dimensional combustion characteristics for methane--oxygen mixtures \cite{Peswani_IMECE2020,Peswani_ussci2021,Peswani2022}.  {While reduced elementary reactions mechanisms have been successfully applied at micro-scale resolutions to study transverse detonations in methane-oxygen mixtures \cite{VIJAYAKUMAR2020_thesis}, for example using only 13 species and 35 reactions, the 4-step model was adopted instead owing to its much lower overhead.  This permitted hundreds of simulations to be performed for a wide range of quiescent pressures and resolutions in a timely manner.}  In this model, we considered only the evolution of global species $R0$, $R1$, $P1$, and $P2$. The equivalent reactant and product groups used can be summarized as 
\begin{equation}
    \begin{cases}
        \mathrm{C}\mathrm{H}_4 + 2\mathrm{O}_2 & \rightarrow R0 \textrm{ \& } R1 \\
        \rightarrow \mathrm{CO_2}+2\mathrm{H_2O} & \rightarrow P1 \\
        \rightleftharpoons \mathrm{CO}+4\mathrm{H}+3\mathrm{O} & \rightarrow P2.
    \end{cases}
\end{equation}
As a result of the above grouping, the NASA coefficients of each group of species were determined by considering the sum of individual specie coefficients multiplied by their mole fraction (in the species group) \cite{Zhu_2012}.  The reaction paths were built by fitting the reference data from constant volume processes using the detailed GRI-3.0 mechanism \cite{GRI} in Cantera \cite{cantera}. Although newer mechanisms have been developed for high pressure C1-C4 combustion \cite{wang_mechanism,Li2017}, the selected GRI-3.0 mechanism was deemed appropriate since the conditions encountered in this study are moderate, up to $T=1500$ K and $p=3.5$ atm in the unburned gas.  This is well within the range of pressures for which the GRI-3.0 mechanism was optimized. The reaction path fitting was done by substituting the global species for a reactive mixture ($R$, $P1$, and $P2$) into the process while conserving the overall thermodynamic properties. The reaction paths, and corresponding reaction rates and orders, were acquired by modeling the reaction as having two thermally neutral induction regime paths, two irreversible exothermic reaction paths that convert $R$ to $P1$ and $P2$ separately, and an additional equilibrium step between $P1$ and $P2$. The reaction scheme can be summarized as
\begin{equation}
    \begin{cases}
      (i1)\hspace{5mm} R0 \rightarrow R1 & k_{i1}\\
      (i2)\hspace{5mm} R0 + s_0\cdot R1 \rightarrow (1+s_0)\cdot R1  & k_{i2}\\
      (r1)\hspace{5mm} R1 \rightarrow \delta_1 \cdot P1 \hspace{5mm}  & k_{r1}\\
      (r2)\hspace{5mm} R1 \rightarrow \delta_2 \cdot P2 \hspace{5mm}   & k_{r2}\\
      (e)\hspace{5mm} P1 \rightleftharpoons \delta_3 P2 \hspace{5mm}   &k_{ef},k_{er},
      \label{reaction_eq}
    \end{cases}\newline
\end{equation}
where the absolute reaction rate constants $k_{i1}, k_{i2}, k_{r1}, k_{r2}, k_{ef},$ and $k_{er}$, and reaction order $s_0$, rely only on the local thermal state of the mixture, while the stoichiometry coefficients are:
$\delta_1 = W_R/W_{P1}, \delta_2 = W_R/W_{P2}, \delta_3 =W_{P1}/W_{P2}$, where $W_i$ is the molecular weight of the $i$th species group. The species $R1$ in the model plays the role of an activated reactant meant to replace the numerous radicals and intermediate species that are formed during a typical combustion process from reactants to products.  By making use of the fact that $\sum{Y_i}=1$, only three transport equations are needed in place of Eq.~\eqref{eqn.reactant_euler}.

\begin{figure}[h!]
\centering
\includegraphics[width=88mm]{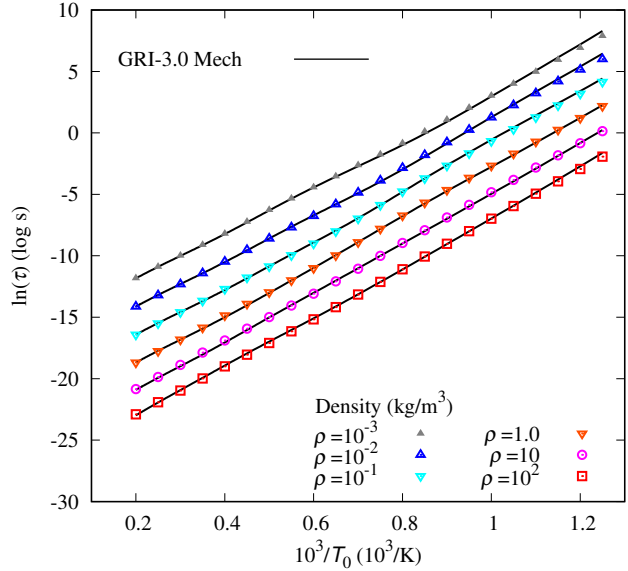}
\caption{Constant volume ignition delay times ($\tau$) for stoichiometric methane--oxygen computed using the four-step model \cite{Peswani_IMECE2020} and compared to the detailed GRI-3.0 mechanism \cite{GRI} for a wide range of initial densities ($\rho$) and temperatures ($T$).}
\label{ignitionTimes}
\end{figure}

\begin{figure}[h!]
\centering
\includegraphics[width=88mm]{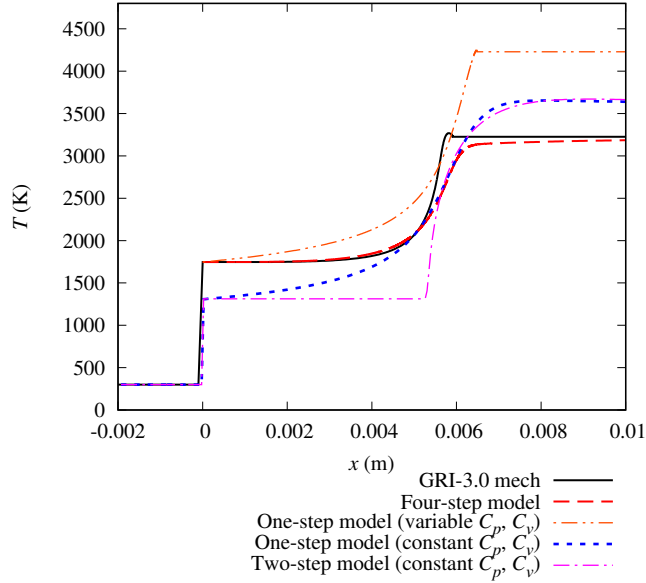}
\caption{{Temperature profiles behind a Mach 6.35 (2262.6 m/s) shock in stoichiometric methane--oxygen at $T_0=300$ K and $p_0=5.5$ kPa computed using conventional one- and two-step perfect gas models \cite{lau2013numerical}, a one-step model with temperature dependent heat capacities, the four-step model \cite{Peswani_IMECE2020}, and the detailed GRI-3.0 mechanism \cite{GRI}.}}
\label{znd_profiles}
\end{figure}

Specific details of the combustion model, including the calibrated parameters and model performance at capturing zero- and one-dimensional combustion problems, are found elsewhere \cite{Zhu_2012,Peswani_IMECE2020,Peswani_ussci2021,Peswani2022}.  However for completeness, Fig.~\ref{ignitionTimes} demonstrates the four-step model ability to capture constant volume ignition delay times for stoichiometric methane--oxygen at a wide range of initial temperatures and pressures when compared to the GRI-3.0 mechanism.  {Also, Fig.~\ref{znd_profiles} compares the temperature profiles obtained behind a Mach 6.35 (2262.6 m/s) shock in stoichiometric methane--oxygen at $T_0=300$ K and $p_0=5.5$ kPa using the four-step model, conventional one- and two-step calorically perfect gas models \cite{lau2013numerical}, a one-step with temperature dependent heat capacities, and the detailed GRI-3.0 mechanism \cite{GRI}.  Here we first note that the conventional one- and two-step perfect gas models \cite{lau2013numerical} do not capture the correct post-shock or post-reaction state variables (i.e., temperature).  Although the induction lengths have been tuned to the conditions behind the given incident shock strength, we point out that such tuning was actually performed at the wrong temperature (and pressure).  Should a second shock form in the shocked mixture, and since the induction lengths were tuned only to the conditions behind the first shock, it is very likely that the ignition time would not be correct since the state variables would deviate further from the detailed chemistry.  We also note that the one-step model performs poorly at minimizing heat release in the induction zone, which of course impacts the local ignition delay times and their gradients behind the shock.  In fact, it was previously demonstrated that temperature gradients
capable of allowing detonations to form calculated using detailed chemical models is much shallower compared to those predicted by simple chemical models \cite{WANG2018400}.  This is likely due to the sensitivity of local ignition delay times and coupling of shock and reaction zones to the temperature of the gas.  Although a one-step combustion model with temperature dependent heat capacities would perform better at capturing the post-shock states, as shown, and could be tuned to reproduce the ignition delays in a wide range of temperatures and pressures, we note the incorrect product state.  In this simple model, we considered only the reaction of $R0\rightarrow P1$, governed by an Arrhenius reaction rate law in the form}
\begin{equation}
    \dfrac{\textrm{D}Y_{\textrm{R0}}}{\textrm{D}t} = A \rho^m T^n  [\textrm{R0}]^{s} \exp{\biggl(\frac{-(Ea/R)}{T}\biggr)}.
\end{equation}
{Here, $A=2\times10^{12}$, $m=0.2$, $n=-0.6$, and $(Ea/R)=20,562$ K was used. In this model, equilibrium with products forming species $P2$ was not considered, yet the formation of such incomplete combustion products are known to be heavily dependent on the state variables and also highly influential on the final enthalpy obtained \cite{Zhu_2012}. This shortcoming would likely lead to incorrect detonation velocities, since the enthalpy change (or heat release) thus differs significantly from the detailed chemistry.  The four-step model used, on the other hand, is a minimal global combustion model with equilibrium effects that is able to reproduce the detailed detonation structure, with the exception of minor departures in the reaction zone stiffness, as shown in the figure.}

Finally, to solve the governing equations, Eqs.~\eqref{eqn.mass_euler}-\eqref{eqn.reactant_euler}, the second order HLLC method \cite{Toro2009} was applied, using the van Albada slope limiter \cite{vanAlbada1982}.  The usual operator splitting approach was applied, where the hydrodynamic evolution was solved first using a CFL number of 0.4, followed by adding the first order source term evaluation across the same time-step.  The source terms ($\dot{\omega}_i$) were evaluated using the implicit backward Euler method based on Newton iteration, and implemented using the Sundials CVODE libraries \cite{Hindmarsh2005}.  Adaptive mesh refinement (AMR) \cite{Falle1993} was also applied to compute detailed solutions only in regions of interest, such as the shocked and unburned gas.  For this study, a computational cell was flagged as needing refinement if $Y_{R1}>0.001$, or if $Y_{R0}>0.99$ and $\rho>1.1\rho_0$, where $\rho_0$ is the density of the quiescent fluid.  Cells were also flagged as needing refinement when density changes of more than 10\% occurred between grid levels.  Finally, the grid was always refined along the boundary of the internal half-cylinder geometry.  When a cell was flagged as `bad', or needing refinement, the badness was also diffused by a few cells to ensure smooth solutions across fine-course cell boundaries. The base grid resolution for all cases was 10 mm, with anywhere between 4 to 8 additional levels of refinement applied, depending on the desired minimum grid resolution.

\subsection{Domain, initial and boundary conditions}
\begin{figure}[h!]
\centering
\includegraphics[width=88mm]{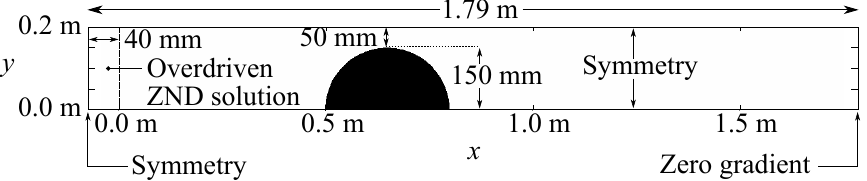}
\caption{Numerical set-up, with a zero gradient condition on the right boundary, and symmetric elsewhere (including the obstacle).}
\label{domain}
\end{figure}

This study simulated stoichiometric methane-oxygen detonation interactions with a half-cylinder obstacle, corresponding to the past experiments \cite{Bhattacharjee2013b,Bhattacharjee2013}.  A 150mm radius half-cylinder was modeled in a two-dimensional channel of 200 mm height and 1.75 m length, as shown in Fig.~\ref{domain}.  An initially overdriven Zel'dovich-von Neumann-Doring (ZND) solution was imposed at $x=0$ that was oriented to propagate to the right in the positive $x-$direction, while the left boundary is placed at $x=-40$ mm. In order to overcome startup errors associated with sharp discontinuities, the ZND solution was given an overdrive factor ($f$) of 1.2, where $f=(U_\mathrm{s}/U_\mathrm{CJ})^2$.  Here, $U_\mathrm{s}$ is the overdriven shock speed, while $U_\mathrm{CJ}$ corresponds to the CJ-detonation speed.  The right boundary condition is a zero-gradient type, while the remaining boundaries, including the cylinder surface, are symmetric type in which only the normal velocity components are reversed.  {The left boundary condition thus deliberately creates a Taylor-Wave structure \cite{Taylor1950}, whose intention is to slow the overdriven wave down to the CJ-speed prior to its interaction with the cylinder.  Once the CJ-speed is reached, the flow of products becomes choked.  Beyond this, the expansion wave has no effect on the detonation wave front.}  The leading edge of the cylinder is placed 500 mm from the initial ZND wave.  This distance was found to be sufficiently long to permit the detonation wave to settle to within 3\% of the CJ-detonation speed by the time the wave reached the throat of the cylinder, for all initial pressures and resolutions considered.  {Finally, the cylinder surface was treated using a conventional \emph{staircase} approach.  Implications of this approach, including our justification for its adoption is discussed in \ref{sec.InternalGeometry}}.  In order to observe the different regimes expected (detonation quenching, critical ignition, critical detonation re-initiation, and transmission), the initial pressure was varied anywhere from $p_0=3.5$ kPa to 16 kPa.  This choice of pressure ranges was based on the experimental results of Bhattacharjee \cite{Bhattacharjee2013b}.   The initial temperature for all simulations was $T_0=300$ K.

\section{Results}\label{sec:Results}

\subsection{Regimes observed at the first shock reflection}

In this section, we present an overview of the several different outcomes observed that resulted from varying the initial pressure.  The minimum resolution used here was 78 {\gmu}m, which was found sufficient to capture different regimes and flow features of interest.  At this resolution, 21.6 to 55.7 grids per ZND induction lengths (or 4.7 to 11.9 grids per reaction lengths) were captured, depending on the initial quiescent pressure. This resolution is also consistent with past numerical investigations that used one- and two-step combustion modeling \cite{Bhattacharjee2013,lau2013numerical}.     The effects of grid resolution is presented in Sec.~\hyperref[sec:grid]{3.2}.

\begin{figure}[h]
	\begin{center}
		\includegraphics[width=88mm]{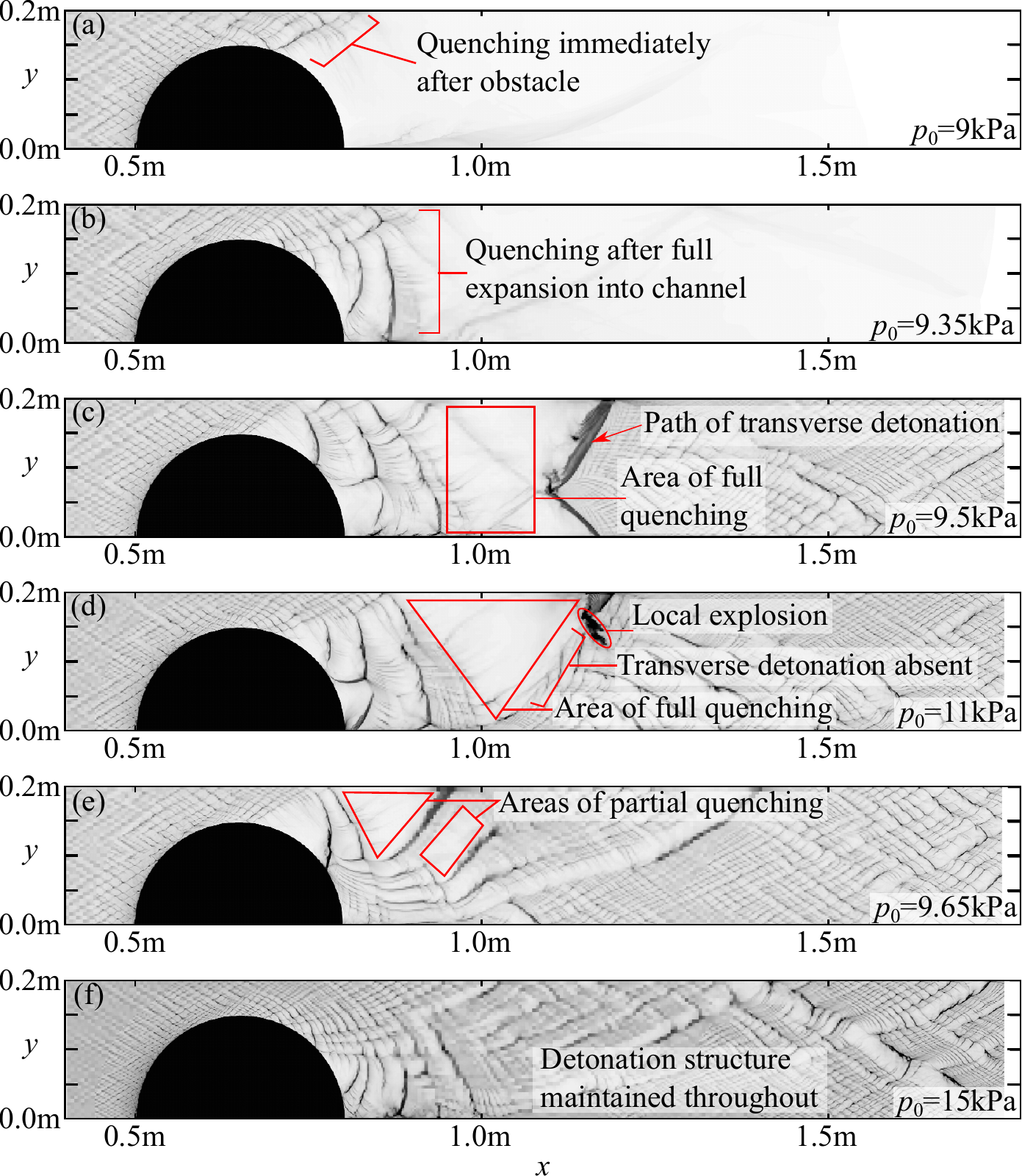}
	\end{center}
	\caption{Numerical soot foils showing the six detonation behavior regimes, all at 78 {$\upmu$}m resolution: (a) detonation quenching, (b) critical ignition without detonation, (c) critical detonation re-initiation, (d) critical detonation re-initiation without transverse detonation, (e) critical transmission, and (f) unattenuated detonation transmission.}
	\label{fig.class_sf}
\end{figure}

In general, six different possible outcomes were observed, and were found to somewhat depend on the initial pressure.  To classify each case, both qualitative and quantitative information was used to determine the category of behavior seen in each simulation. Figure \ref{fig.class_sf} shows example numerical soot foil images obtained for each of the six regimes, with the main distinctive features labeled. These numerical soot foil images were obtained by recording the maximum pressure ever experienced locally in each grid point throughout the course of the simulation. Since the pressures are highest at triple point locations, where incident, Mach, and transverse shocks meet, this method effectively tracks the trajectories of triple points.  Figure \ref{fig.class_sf}a displays \emph{detonation quenching} (DQ) for $p_0=9$ kPa, where the detonation completely quenched immediately after clearing the obstacle. A detonation is quenched when the cellular pattern disappears, signaling a decoupling of the shock front and reaction zone that is never re-established into a detonation. Figure \ref{fig.class_sf}b shows \emph{critical ignition} (CI) without detonation re-initiation.  In this case, significant burning of the reactants occurred behind the Mach shock, but a reaction zone did not couple to the shock. This can be seen in the soot foil image as the detonation took longer to fully quench. Figure \ref{fig.class_sf}c shows the main regime of interest in this study, \emph{critical detonation re-initiation} (CDR). This is characterized by an area of complete quenching, followed by re-initiation that features one or more transverse detonations.  In the particular case shown, the transverse detonations are characterized by dark bands that started at the center and propagated toward the upper and lower boundaries of the channel. In other cases, a transverse detonation started near the lower boundary and propagated upwards. Figure \ref{fig.class_sf}d is \emph{critical detonation re-initiation without transverse detonation} (CDR-NTD), which also had an area of complete quenching, but a transverse detonation was not observed before the detonation became fully established again. It is important to note that this regime was a less frequent outcome compared to CDR, numerically, except at the lowest resolution of 625 {\gmu}m.  We also note that Bhattacharjee reported the experimental outcome to be rare, and could not be easily reproduced \cite{Bhattacharjee2013b}. In the simulation shown, a localized explosion occurred near the top boundary, but this was not the mechanism that triggered detonation re-initiation along the Mach shock front. Figure \ref{fig.class_sf}e shows \emph{critical transmission}, where partial quenching occurred after interaction with the cylinder. Partial quenching is characterized by the presence of areas without the cellular structure, but the quenched areas never spanned the entire height of the channel.  Finally, Fig.~\ref{fig.class_sf}f displays \emph{unattenuated detonation transmission}, where no significant areas of quenching were observed, and the cellular structure was maintained throughout the length and height of the channel. 

\begin{figure}
	\begin{center}
		\includegraphics[width=88mm]{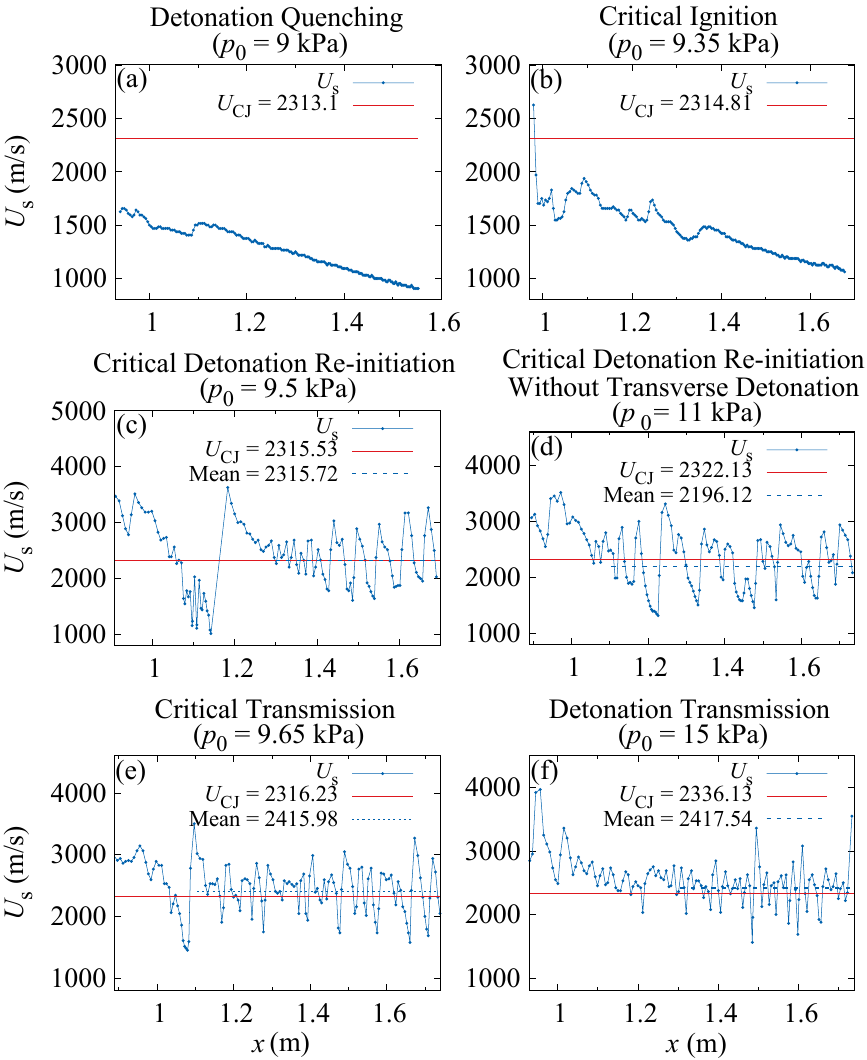}
	\end{center}
	\caption{Speed of the wave front ($U_\mathrm{s}$) as a function of $x$-distance along the channel length for each detonation behavior regime. All measurements are taken on the bottom boundary, and all simulations were conducted with a minimum grid resolution of 78 {$\upmu$}m. Averages were only taken for regions of quasi-steady detonation propagation.  The end of the obstacle is located at $x = 0.8$ m.}
	\label{fig.class_speed}
\end{figure}

Figure \ref{fig.class_speed} shows the speed of the leading shock wave ($U_\mathrm{s}$) vs.~$x-$position along the channel, obtained along the bottom boundary at $y=0$. There are clear differences observed between each case. Figure \ref{fig.class_speed}a shows detonation quenching, where the speed of the leading shock wave along the bottom boundary after shock reflection never reached the theoretical CJ speed of 2313.1 m/s; the wave speed continually decreased with distance. Critical ignition without detonation, shown in \ref{fig.class_speed}b, had a similar outcome.  However, in this case the leading shock speed started $\sim11$\% above $U_\mathrm{CJ}$, and then decayed rapidly to $U_\mathrm{s}\approx 1700$ m/s (26.6\% below $U_\mathrm{CJ}=2314.81$ m/s). From $1\le x\le 1.4$ m, fluctuations were observed in the wave speed.  Beyond $x>1.4$ m, the leading shock speed continually decayed with distance. Critical detonation re-initiation is shown in \ref{fig.class_speed}c, where the wave began $\sim49$\% above $U_\mathrm{CJ}$ after encountering the obstacle, and then gradually decayed for $0.95\le x\le 1.15$ m to $\sim56$\% below the CJ value of 2315.59 m/s. At the re-initiation event (at $x \approx 1.15$ m), there was a sudden dramatic increase in speed to 3625.0 m/s.  Beyond this distance, the detonation eventually settled to and oscillated around the CJ speed once again, with an average speed of 2315.72 m/s, $\sim.008$\% above the CJ value. Figure \ref{fig.class_speed}d displays critical detonation re-initiation without transverse detonation where the wave speed began $\sim32$\% above the CJ value of 2322.13 m/s due to the shock reflection and then gradually decayed. The process continued with periodic triple point collisions, leading to consistent oscillations of speed, with an average of $U_\mathrm{s}=2196.12$ m/s, $\sim5$\% below $U_\mathrm{CJ}$. Critical transmission (Fig.~\ref{fig.class_speed}e) also displayed gradual speed decay for $0.95\le x\le 1.1$ m, starting from an overdriven speed 28.17\% over the CJ speed of 2415.98 m/s.  The speed history of this case is very similar to both Figs.~\ref{fig.class_speed}c and d, which explains the necessity to pair the speed classification with qualitative observations in order to get a clear picture of the overall combustion regime behavior. Unattenuated detonation transmission (Fig.~\ref{fig.class_speed}f) began overdriven (up to $\sim69.9$\% above $U_\mathrm{CJ}$) and then decayed to an average propagation speed of 2417.54 m/s, $\sim3.5$\% above $U_\mathrm{CJ}$. The initially high values for speed for the latter five cases are due to a shock reflection along the bottom boundary.

\subsubsection{Detonation quenching}

\begin{figure}[H]
\centering
\includegraphics[width=88mm]{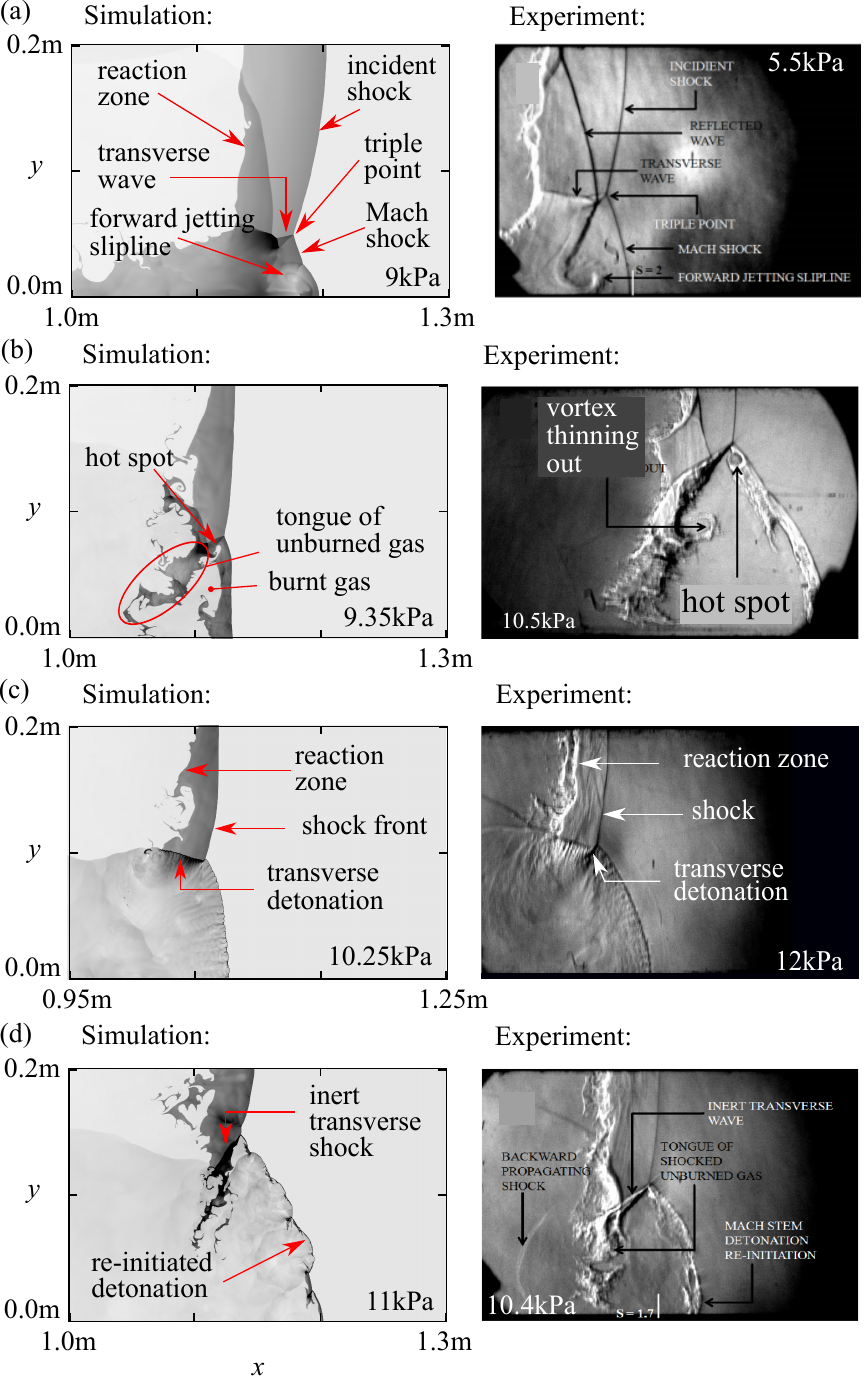}
\caption{Different simulated density fields at 78 {$\upmu$}m resolution (left column) and compared to experimental schlieren images, reproduced from R.~Bhattacharjee \cite{Bhattacharjee2013b} with permission of the author (right column): (a) detonation quenching, (b) critical ignition without detonation, (c) critical detonation re-initiation, (d) critical detonation re-initiation without transverse detonation. The height of each image is 200 mm.}
\label{cases_summary}
\end{figure}

\begin{figure}[h]
\centering
\includegraphics[width=88mm]{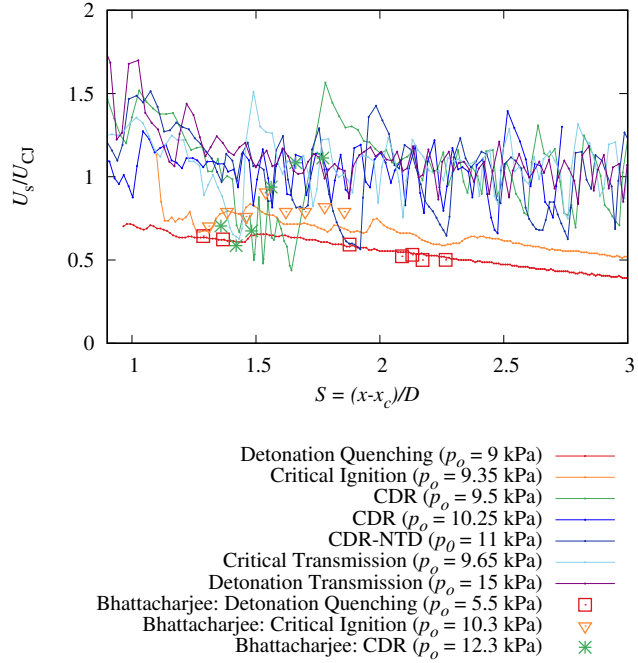}
\caption{Speed of the Mach shock along the bottom boundary normalized by the CJ detonation speed ($U_\mathrm{s}/U_\mathrm{CJ}$) vs.~normalized x-position ($S=(x-x_c)/D$) for different initial pressures ($p_0$) and minimum grid resolution of 78 {$\upmu$}m. Critical Detonation Re-Initiation is denoted as CDR.}
\label{fig:speed_bottom}
\end{figure}

In Fig.~\ref{cases_summary}a, we show that at sufficiently low initial pressures, for example at $p_0=9$ kPa, the simulated density field of the fully quenched detonation wave resulting from the diffraction around the obstacle compares well qualitatively to the schlieren photograph of a past experiment \cite{Bhattacharjee2013b}.  The figure shows a clear separation between the shock front and reaction zone. This separation occurs because the detonation front is allowed to expand after the obstacle, leading to an increased surface area of the shock. The increased area leads to a weakened shock strength, which lengthens the ignition delay times and therefore increases the distance between the shock front and reaction zone \cite{arienti_shepherd_2005}. The various shock dynamics, including the incident shock, Mach shock, transverse, and reflected waves are all captured well compared to the experiment.  The slip line and forward jetting are also captured.  We note that although the experimental image was captured at a much lower pressure ($p_0=5.5$ kPa), few experiments were conducted in the range of $p_0=6$ to 10 kPa, and thus an exact quenching limit was not found experimentally.  Quantitatively, the normalized Mach shock speed on the bottom wall ($U_\mathrm{s}/U_\mathrm{CJ}$) was found to compare favorably to values measured experimentally by Bhattacharjee \cite{Bhattacharjee2013}, as shown in Fig.~\ref{fig:speed_bottom}. In this figure, the abscissa is the normalized distance from the center of the cylinder, characterized by $S=(x-x_c)/D$, where $x_c$ is the cylinder center location and $D$ is the cylinder diameter.

From our simulations, failure always occurred below $p_0\le 8$ kPa at the 78 {$\upmu$}m resolution, and sometimes up to $p_0=10.5$ kPa.  In this regard, the outcome appears to be stochastic in this pressure range. By measuring the detonation cell size just prior to its interaction with the cylinder, using an autocorrelation procedure~\cite{Sharpe2011}, we find that the limit to ensure failure, at this resolution, is {$(d_\mathrm{H}/\lambda)_\textrm{fail}<4.3$}, where $d_\mathrm{H}$ is the size of the gap between the cylinder and top boundary, and $\lambda$ is the cell size at evaluated for $p_0\le8$ kPa.  This result is about 5 to 10 times greater than Bhattacharjee's result, where $(d_\mathrm{H}/\lambda)_\textrm{fail}=0.5$ to 1.0 \cite{Bhattacharjee2013b}. Since experiments were not documented between $p_0=6$ to 10 kPa, it is likely that the actual limit for $(d_\mathrm{H}/\lambda)_\textrm{fail}$ could be higher than reported experimentally.

\subsubsection{Critical ignition without detonation}

In Fig.~\ref{cases_summary}b, at a slightly elevated pressure of $p_0=9.35$ kPa, a critical ignition regime was observed where significant burning occurred behind the Mach shock, yet the detonation front did not re-establish.  Speed measurements of the Mach shock along the bottom boundary, shown in Fig.~\ref{fig:speed_bottom}, revealed that the wave traveled faster than the fully quenched case.  However, both cases (CI and DQ) traveled slower than the CJ-detonation speed, and both cases continually decelerated with distance away from the obstacle.  While the normalized simulated Mach shock traveled below the experimentally reported values (by $\sim6.8$\%), the general behavior of the wave front was consistent.  In this case, an ignition hot spot immediately behind the newly formed Mach shock coupled to forward jetting and Kelvin-Helmholtz instabilities along the slip line lead to enhanced combustion during the evolution of the Mach shock.  Additional hot spots formed as indicated in the figure, and a \emph{tongue} of unburned gas formed behind the slip line.  The outcome is qualitatively similar to the experimental observations, however a closer coupling of shock and reaction zone was observed in the experiments \cite{Bhattacharjee2013}.  The experiment showed nearly all of the gas behind the Mach shock as burned, whereas the simulation had a smaller area of burned gas. We note here that previous numerical modeling of the critical ignition case \citep{Maxwell2016_thesis} has highlighted the importance of providing closure to turbulent diffusion in order to properly capture forward jetting and mixing of combustion reactants and products in this critical ignition regime. At this time, we have not provided such closure.

\subsubsection{Critical detonation re-initiation}

In the range $8.5\le p_0 \le 13$ kPa, different critical outcomes were observed where fully quenched detonations re-initiated after the first shock reflection with the bottom boundary.  Figure~\ref{cases_summary}c shows a typical critical detonation re-initiation outcome at $p_0=10.25$ kPa where a sustained detonation was re-initiated along both the Mach shock and also along the transverse wave.  In Fig.~\ref{fig:speed_bottom}, the Mach shock speed relative to the CJ speed ($U_\mathrm{s}/U_\mathrm{CJ}$) vs.~normalized distance from the cylinder center ($S$) is shown for two CDR cases.  For $p_0=9.5$, the Mach shock was initially traveling at a speed below the CJ-detonation speed and with a decelerating trend, following closely the critical ignition case without detonation re-initiation at $p_0=10.3$ kPa.  However, after some distance, around $S\approx1.7$ m, the wave momentarily became overdriven, up to $(U_\mathrm{s}/U_\mathrm{CJ})=1.56$.  For $p_0=10.25$, the wave speed was initially below the CJ value, at $(U_\mathrm{s}/U_\mathrm{CJ})=0.88$, but became overdriven to $(U_\mathrm{s}/U_\mathrm{CJ})=1.27$ much sooner, at  $S=1.04$.  For both cases, the wave eventually settled to an average value around the CJ-detonation speed as the quasi-steady detonation wave was re-established.  For comparison, experimental normalized Mach shock speed measurements are also shown in Fig.~\ref{fig:speed_bottom} for $p_0=12.3$ kPa, which also had an overdriven Mach shock speed of $(U_\mathrm{s}/U_\mathrm{CJ})_\mathrm{exp}=1.09$ immediately after re-initiation.  

We draw attention to this particular simulated critical outcome, as the sustained transverse detonation feature was never observed in past numerical attempts at modeling this scenario \cite{Radulescu2011,Bhattacharjee2013,lau2013numerical,MAXWELL2018340}.  We attribute this success to the adoption of the thermally perfect four-step combustion model, which was calibrated to reproduce the the correct ignition delays at different temperatures and pressures when compared to the GRI-3.0 mechanism \cite{GRI}.  It is worth noting that while past numerical simulations have been successful in capturing transverse detonation waves in critical diameter problems involving hydrogen \cite{GALLIER20172781,mevel2017hydrogen} and highly irregular near-limit detonation propagation in methane--oxygen \cite{VIJAYAKUMAR2020_thesis}, such simulations have generally needed to use detailed elementary combustion mechanisms.   Exceptions to this are highly irregular critical detonation propagation \cite{gamezo2000marginal} and critical detonation diffraction \cite{shi_uy_wen_2020} where the transverse detonations were in fact observed using one-step models. However, the combustion models in these cases required calibration of the heat release to force the desired unstable detonation behavior, and did not necessarily reproduce all of the combustion characteristics of a particular reactive mixture.

\subsubsection{Critical detonation re-initiation without transverse detonation}

\begin{figure}
	\begin{center}
		\includegraphics[width=88mm]{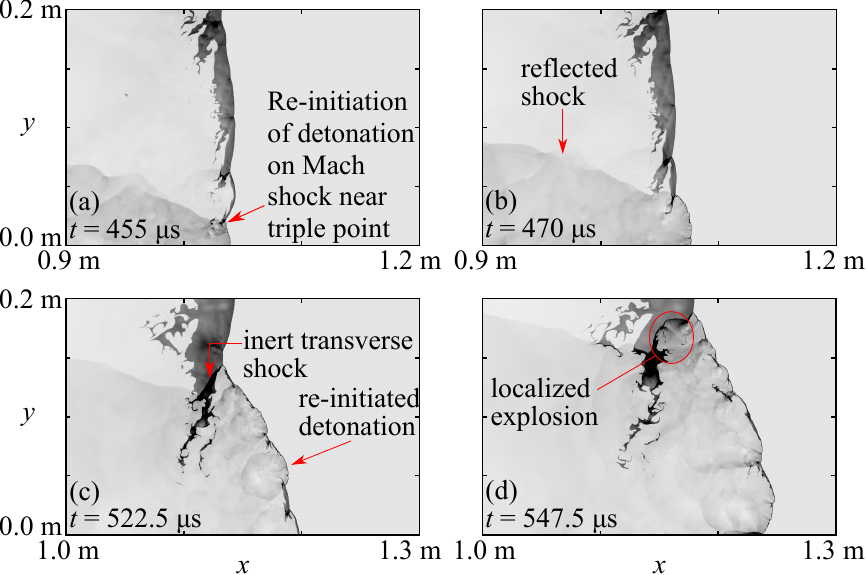}
	\end{center}
	\caption{Density evolution of a simulated critical detonation re-initiation without transverse detonation outcome for an initial pressure of $p_0 = 11$ kPa.}
	\label{fig.NTD}
\end{figure}

At $p_0=10$ and 11 kPa, critical cases were observed where detonation re-initiation was only observed along the Mach stem.  This is shown for the $p_0=11$ kPa case in Fig.~\ref{cases_summary}d.   A similar set of features in the flow field between simulation and experiment was observed for this case. The Mach stem re-initiated into a detonation, accompanied by an inert shock that propagated upward, and a tongue of shocked unburned gas that trailed behind. In both the simulation and experiment, the transverse wave remained inert, with the tongue of unburned gas consumed only through deflagration.  The sequence of events for the case shown in Fig.~\ref{cases_summary}d is presented in more detail in Fig.~\ref{fig.NTD}. Once the detonation had quenched, the detonation was re-initiated directly on the Mach shock near the triple point, as shown Fig.~\ref{fig.NTD}a, yet no transverse detonation was present. The newly established detonation front continued to expand (Fig.~\ref{fig.NTD}b), and eventually the inert transverse shock became apparent (Fig.~\ref{fig.NTD}c). Once the triple point had almost reached the entire height of the channel, a localized explosion occurred on the inert transverse shock as shown in Fig.~\ref{fig.NTD}d.  This localized explosion was responsible for the dark bands previously shown in the numerical soot foil image of Fig.~\ref{fig.class_sf}d.

It is important to note here that in the past numerical investigations based on calorically perfect one- or two-step model approaches \cite{Radulescu2011,Bhattacharjee2013,lau2013numerical,MAXWELL2018340}, this mechanism of detonation re-initiation was almost always observed. In this numerical investigation, however, and also in the experiments \cite{Bhattacharjee2013b}, this outcome was observed less frequently.  In fact, most critical detonation re-initiation cases, regardless of specific scenario considered, have long been speculated to occur with the presence of a transverse detonation \cite{Chao2006,Radulescu2011,Edwards1979,ohyagi2002diffraction}. Thus, the physical observance of this specific outcome, CDR without transverse detonation, is not typical and also appears to be stochastic. When the past experiment was repeated, with the same initial pressure, different results were obtained, and that the observed CDR without transverse detonation outcome was not reproducible \cite{Bhattacharjee2013b}. Likewise, when this simulation was conducted again (at $p_0 = 10$ kPa), different behavior was observed (i.e.~critical transmission), which reveals how sensitive the regimes are to the state of the cellular structure in the pore prior to the diffraction process.

\subsubsection{Critical transmission}

\begin{figure}
	\begin{center}
		\includegraphics[width=88mm]{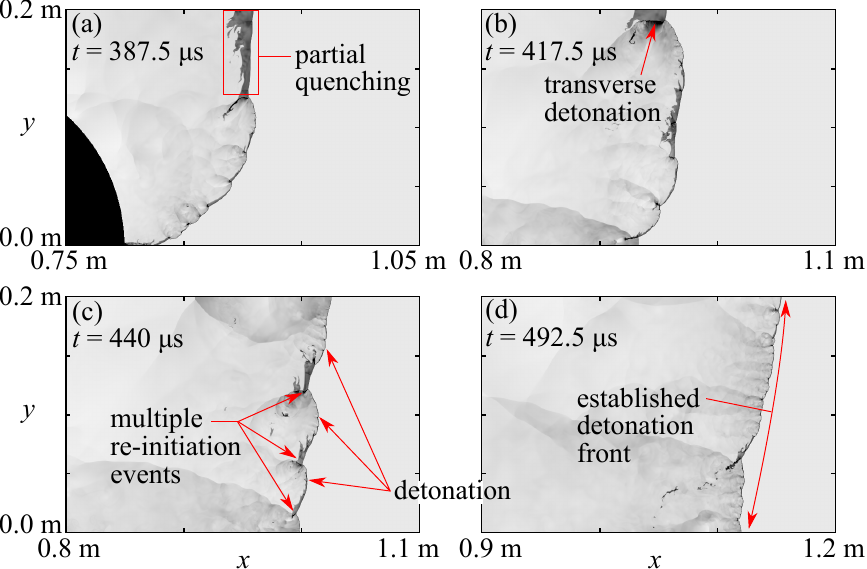}
	\end{center}
	\caption{Density evolution of a simulated critical transmission outcome for an initial pressure of $p_0 = 9.65$ kPa.}
	\label{fig.ct}
\end{figure}

Critical transmission is a regime that was not observed in the past experiments. This could be attributed to the lack of soot foils and sufficient number of diagnostic schlieren images in the experiments.  In the experiments, the observation window was limited to a single location, so the broader perspective was likely missed. This case is comparable to the critical diameter problem \cite{Edwards1979}, where the wave can quench locally but re-initiate on its own before interaction with a shock reflection from the wall boundary. Critical transmission is therefore a newly observed category for this specific scenario being investigated. This regime is characterized by partial quenching of the detonation, where only a portion of the wave front displayed decoupling of the shock front and reaction zone. \emph{Partial quenching} is defined here as a situation where local segments of the quenched detonation wave never span the entire height of the channel. Figure \ref{fig.ct} shows frames of the density field from the $p_0 = 9.65$ kPa case. The first quenching event occurred in Fig.~\ref{fig.ct}a, and it only spanned the top half of the domain height. Figure \ref{fig.ct}b shows that a transverse detonation did form, and traveled through the quenched segment. This largely resembles the process for detonations surviving the critical diameter problem \cite{Pintgen2009655,NAGURA20131949}, and there is likely a strong relation of the origins of the transverse detonations to this case as well. As the wave front propagated forward, there were more quenching and re-initiation events that took place, as shown in Fig.~\ref{fig.ct}c, but there were still portions of the wave that maintained the detonation structure.  This is further supported by the soot foil image from Fig.~\ref{fig.class_sf}e. Eventually, the wave settled into a fully established detonation wave (Fig.~\ref{fig.ct}d) and maintained that state as it propagated through the remaining length of the channel.

\subsubsection{Unattenuated detonation transmission}

Finally, at sufficiently high pressures, i.e.~for $p_0 \geq 10$ kPa at 78 {\gmu}m resolution, unattenuated detonation transmission was observed without any local quenching during the diffraction phase.  After encountering the half-cylinder obstacle, the detonation structure was minimally affected, resulting only in some variation of cell size, i.e.~an increase in average size compared to the structure before the obstacle interaction (refer to Fig.~\ref{fig.class_sf}f).

\subsection{Effects of grid resolution}
\label{sec:grid}

In order to fully interpret the simulation results obtained in this study, it was necessary to perform a grid resolution study in order to understand the influence of changes in grid resolution.  This was especially important since Euler simulations involving detonations are well known to give different solutions with changes in resolution \cite{Radulescu2011,sharpe2001transverse}.   In Euler simulations, deflagrative burning at the interface of the burned and unburned gas can only occur through numerical diffusion. Since a finer resolution results in decreased numerical diffusion \cite{radulescu_sharpe_law_lee_2007}, the laminar burning rates also decrease.  At the same time, turbulent motions are damped at coarser resolutions due to increased numerical diffusion.  In general, multiple grids per detonation induction and reaction lengths are required, so that the details of the reactive hydrodynamic structures may be captured.

\begin{figure}
	\begin{center}
		\includegraphics[width=88mm]{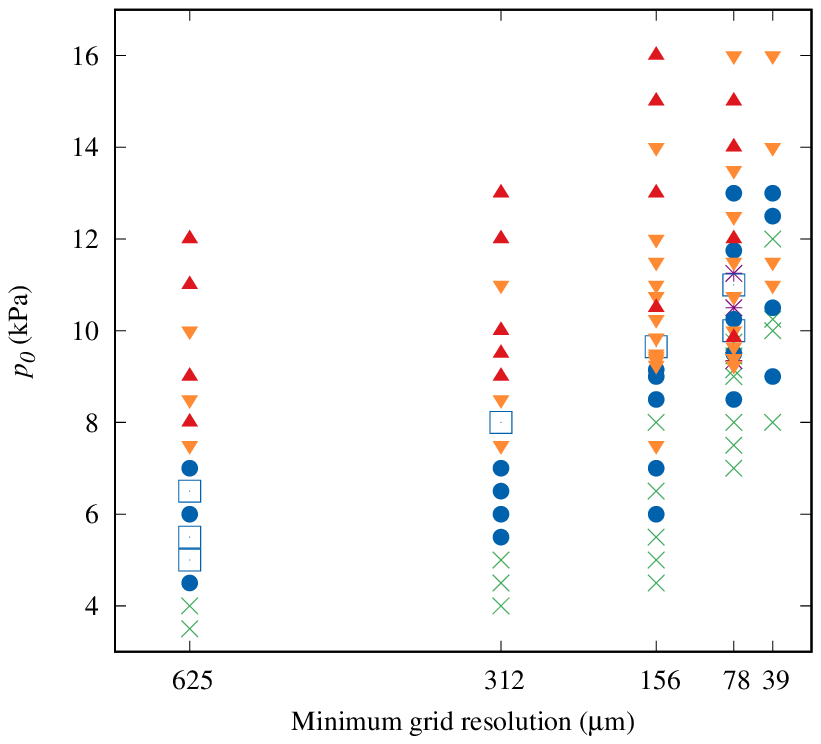}
		\includegraphics[width=88mm]{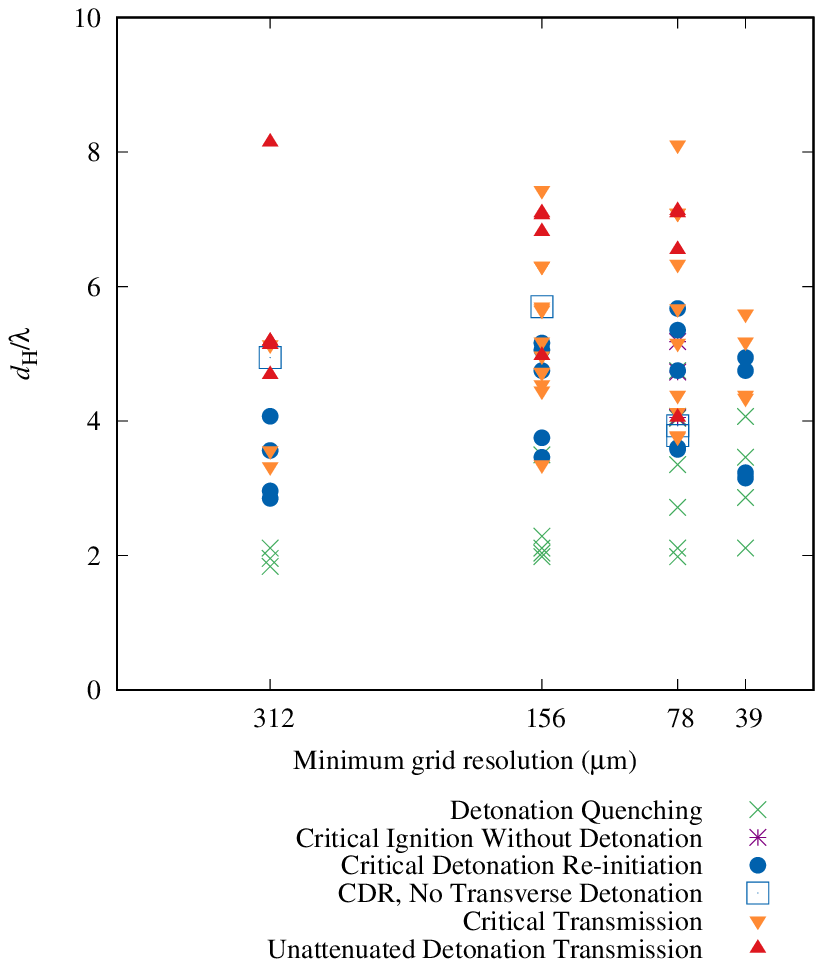}
	\end{center}
	\caption{The different critical regimes observed at each minimum grid resolution and initial quiescent pressure, $p_0$ {(top), and ratio of gap size to mixture cell size, $(d_\mathrm{H}/\lambda)$ (bottom)}.}
	\label{fig.res}
\end{figure}

In this study, simulations were conducted at resolutions as coarse as 625 {$\upmu$}m ($\sim3.8$ to 15.6 grids per induction length) and as fine as 39 {$\upmu$}m ({$\sim50.3$} to 95.6 grids per induction length).  A visual summary of the outcomes at each resolution and initial pressure is shown in Fig.~\ref{fig.res} {(top)}.  In total, more than one hundred simulations were conducted.  It was observed early on that as the resolution becomes finer, the range of pressures encompassing the six categories of behavior shifts upward to higher pressures. For example, the range of pressures that encompass the different regimes in between detonation quenching and transmission is $4<p_0<7.5$ kPa for the coarsest resolution (625 {$\upmu$}m), and $8<p_0<14$ kPa for the finest resolution (39 {$\upmu$}m). As a result, different regimes can be observed at the same pressure across resolutions. Also, while critical outcomes were observed for the coarsest resolutions, 625 {$\upmu$}m, 312 {$\upmu$}m, and  156 {$\upmu$}m, the range of pressures for these regimes is much lower than what was observed experimentally, where CDR was only observed for $10.4 \leq p_0 \leq 16.8$  kPa \cite{Bhattacharjee2013b}.  Even though the range of critical pressures was higher for the 78 {$\upmu$}m and  39 {$\upmu$}m resolutions, where CDR was observed up to $p_0 = 13$ kPa at both resolutions, the departure from the experimental limit can be attributed to losses in the experiments due to boundary layers and heat conduction through the shock tube walls, which were not accounted for in the simulations. Simulations of the critical diameter problem also exhibit this behavior, where it is generally observed that the critical pressures are lower compared to the corresponding experiments \cite{mevel2017hydrogen}.  The principal result of the resolution study conducted revealed that only the 78 {$\upmu$}m resolution was found to capture all of the possible critical outcomes, and the only resolution to capture critical ignition without detonation. 
Furthermore, we note that at the finest resolution of 39 {$\upmu$}m, critical ignition and CDR without transverse detonation were not observed.  The assumption can therefore be made that the amount of numerical diffusion was too low at the finest resolution to allow for an adequate representation of turbulent deflagrative burning that would normally occur in these regimes (critical ignition and CDR without transverse detonation). The 78 {$\upmu$}m resolution was therefore deemed sufficient for observing the mechanisms involved in the detonation re-initiation phenomenon, and was also coarse enough that computational efficiency was ensured.  The range of pressures for the different critical outcomes ($8.5\leq p_0\leq13$  kPa) was also comparable to the past experiments ($10.4 \leq p_0 \leq 16.8$  kPa).

{Also shown in Fig.~\ref{fig.res} (bottom) are the outcomes at each resolution quantified by the ratio of the gap size (throat) to the measured mixture cell width ($d_\mathrm{H}/\lambda$).  Here, the mixture cell width ($\lambda$) was determined for each simulation by applying the autocorrelation procedure~\cite{Sharpe2011} for a numerical sootfoil window where $0.45\le x \le 0.50$.  We first note that the cell size did not converge for increasing resolution, however this was expected since cell sizes obtained in past numerical Euler simulations of methane--oxygen mixtures also did not converge with resolution \cite{radulescu_sharpe_law_lee_2007}.  In fact, it has been demonstrated that for highly irregular mixtures, closure of subgrid-scale turbulent mixing and combustion is required to resolve the correct cellular structure \cite{maxwell_2017}.  Despite this, the results in Fig.~\ref{fig.res} (bottom) reveal that the outcomes remained grid-insensitive at leading order for the resolutions considered.  At the 78 {$\upmu$}m resolution, the critical range was found to be $3.6 \le (d_\mathrm{H}/\lambda)_\mathrm{crit} \le 5.7$, while at 39 {$\upmu$}m this range was shifted slightly lower to $3.2 \le (d_\mathrm{H}/\lambda)_\mathrm{crit} \le 4.9$.  For all resolutions considered, $3 \le (d_\mathrm{H}/\lambda)_\mathrm{crit} \le 6$ to leading order.  In fact, this outcome resembles that of the detonation diffraction problem in rectangular channels where critical outcomes of detonation survival are observed following the abrupt area expansion for channel width to cell size ratios ($W/\lambda$) in the range of 3 to 10 \cite{benedick1983,jones1996influence,deiterding2011high}. In this problem, however, the presence of the cylinder confinement should permit critical transmission of the detonation at larger characteristic mixture cell sizes compared to cases of abrupt expansion.  For example, the converging side of the obstacle experiences a reflected shock, which acts to decrease the effective cell size across the throat, much like detonation propagation into a converging wedge  \cite{Changming2001,Thomas2002,Fortin2015}, and so the actual effective $(d_\mathrm{H}/\lambda)_\mathrm{crit}$ ratio may in fact be larger.  However, the cell size at the throat is difficult to measure, as the reflected wave may not have reached the top wall by the time the detonation reaches the throat.   Also, on the diverging part, a weaker expansion wave is initially felt by the detonation front.  Eventually, however, the expansion of the full $90^\circ$ turn is felt by the detonation front.  Thus, the diverging obstacle effect, in this case, may be predominantly to delay the quenching or critical initiation distance.  Much like the critical diameter problem, there clearly exists a relationship between the mixture cell size and gap size, and most likely also the cylinder geometry itself.}

\begin{figure}
	\begin{center}
		\includegraphics[width=132mm]{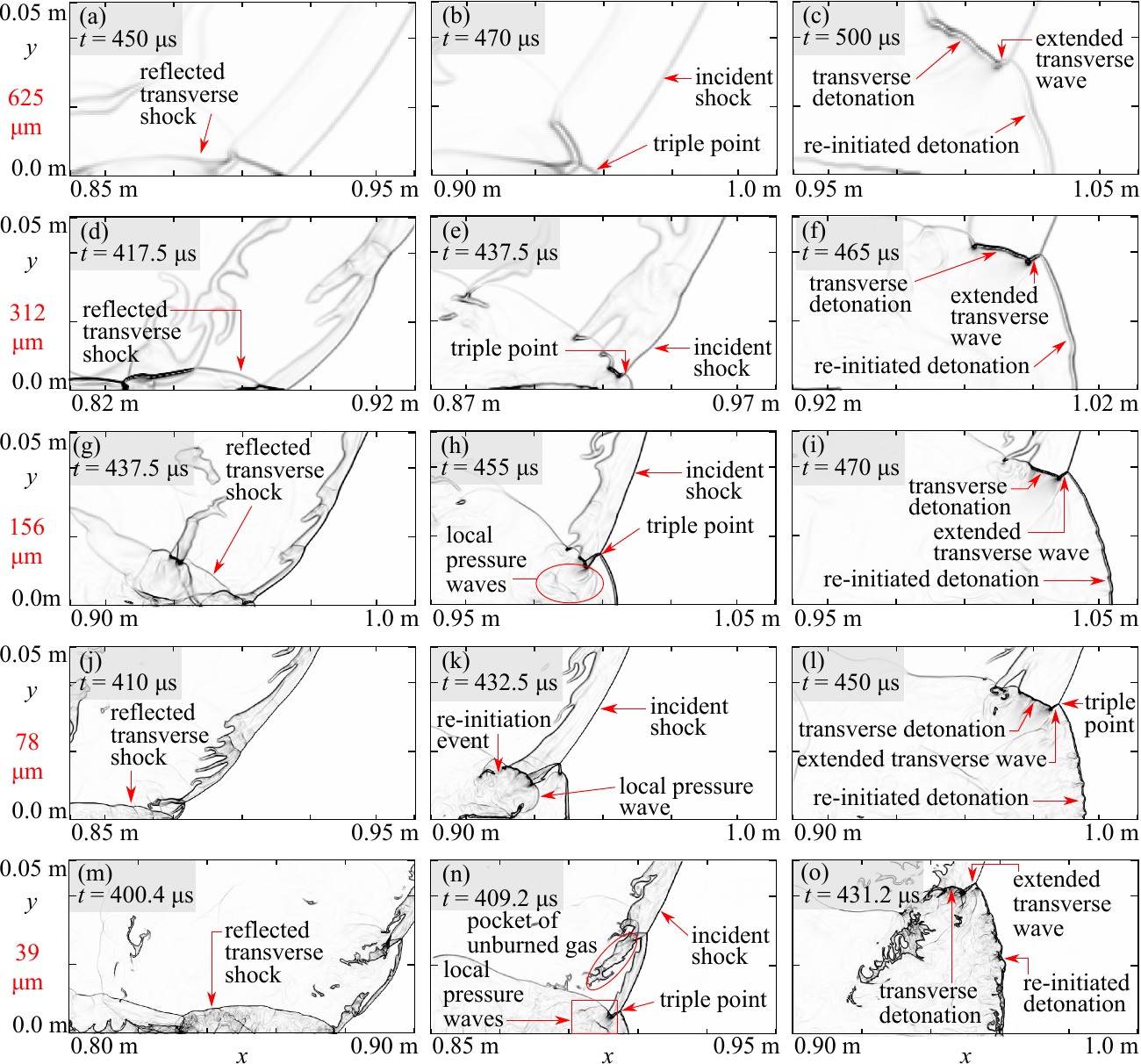}
	\end{center}
	\caption{Density gradient evolution for critical detonation re-initiation as seen at a minimum grid resolution of 625 {$\upmu$}m for $p_0 = 4.5$ kPa (a--c), 312 {$\upmu$}m for $p_0 = 7$ kPa (d--f), 156 {$\upmu$}m for $p_0 = 9.15$ kPa (g--i), 78 {$\upmu$}m for $p_0 = 10.25$ kPa (j--l), and 39 {$\upmu$}m for $p_0 = 10.5$ kPa (m--o).}
	\label{fig.resCDR}
\end{figure}

Despite the differences in regimes observed at different resolutions, and the pressures at which they are observed, the occurrence of the CDR regime are qualitatively similar across resolutions. Figure \ref{fig.resCDR} shows a comparison of the density gradient evolution obtained for CDR outcomes at the six different resolutions tested: $p_0 = 4.5$ kPa at 625 {$\upmu$}m resolution (frames a--c), $p_0 = 7$ kPa at at 312 {$\upmu$}m (frames d--f), $p_0 = 9.15$ kPa at 156 {$\upmu$}m (frames g--i), $p_0 = 10.25$ kPa at 78 {$\upmu$}m (frames j--l), and $p_0 = 10.5$ kPa at 39 {$\upmu$}m (frames m--o). As the resolution becomes finer, more details of the various features present become visible. All resolutions include the key features of a reflected transverse shock, incident shock, triple point, transverse detonation wave, and extended transverse wave. This extended wave is an oblique shock wave and reacting slip line that connects the triple point to the transverse detonation wave. It is a feature that was not explicitly discussed in Bhattacharjee's thesis due to a lack of available resolution in the experimental schlieren photographs \cite{Bhattacharjee2013b}. This feature, however, has been captured numerically before using skeletal detailed elementary reaction mechanisms for Mach stem detonation re-initiation in critical detonation propagation of stoichiometric methane--oxygen \cite{VIJAYAKUMAR2020_thesis} and detonation initiation arising from a double Mach shock reflection in propane--oxygen \cite{ziegler2012simulations}.  In all resolutions, a shock reflection or local explosion drove local pressure waves outward.  In most cases, {the coupling of these pressure waves to the rapid energy release due to chemical reaction led to the initiation of the transverse detonation wave first,} and then the detonation along the Mach stem. Sometimes these {reaction} waves originated from a localized explosion event (see Fig.~\ref{fig.resCDR}k), but in other cases the pressure waves formed directly due to auto-ignition behind the shock reflection itself. The consistency of features observed across resolutions validates the strategy adopted to investigate detonation re-initiation when a transverse detonation is present. The main differences between resolutions were the pressures at which the CDR outcome occurred.  Also there was a prominent pocket of unburned gas present in the finest resolution, seen in Fig.~\ref{fig.resCDR}n, which was not present at the coarsest resolutions.  This can likely be explained by the presence of higher numerical diffusion at coarser resolutions, which lead to quicker burning rates of shocked and unburned gas.

\section{Discussion}\label{sec:Discussion}

\subsection{Origins and the role of the transverse detonations during re-initiation}

Critical detonation re-initiation cases involving a transverse detonation were observed for initial pressures ranging from $8.5\le p_0 \le 13$ kPa at the 78 {$\upmu$}m resolution. Although all other possible cases were observed with some random occurrence in this pressure range, this behavior is consistent with experimental observations of Bhattacharjee \cite{Bhattacharjee2013b}, who noted the stochastic nature of outcomes at critical pressures.

\begin{figure}
	\begin{center}
		\includegraphics[width=88mm]{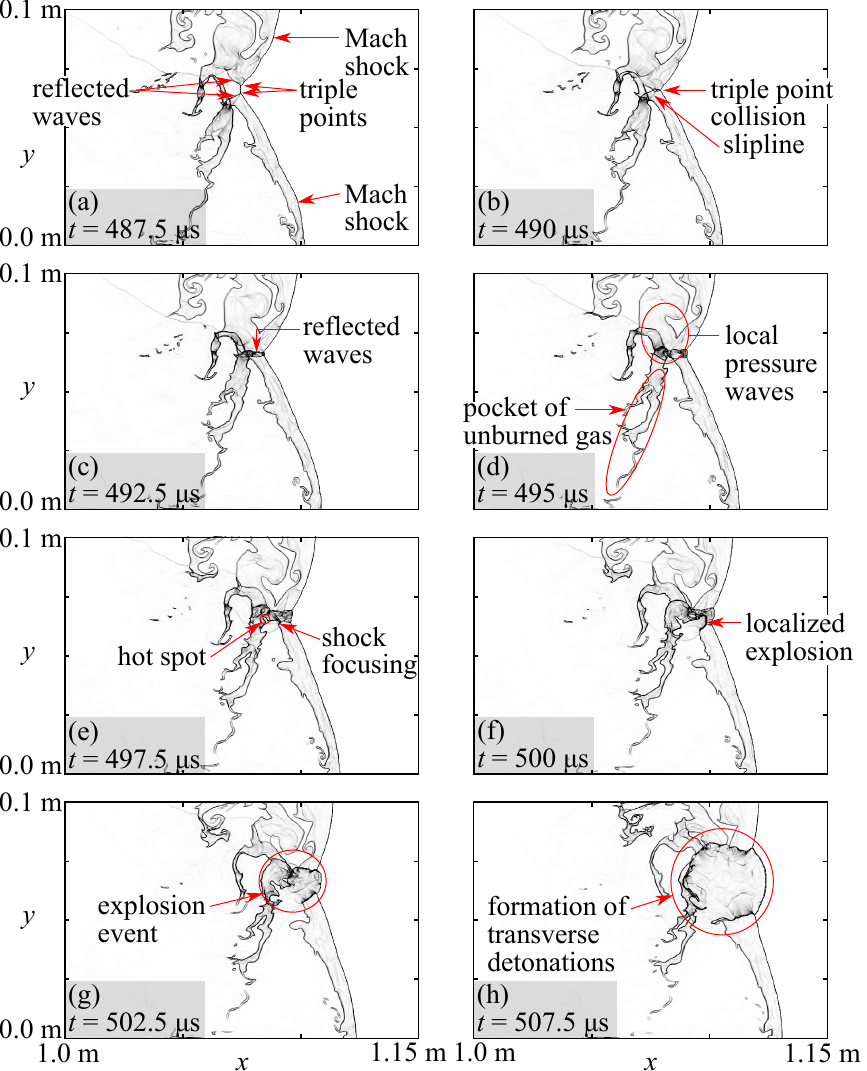}
	\end{center}
	\caption{Density gradient evolution at the moment of detonation re-initiation for an initial pressure of $p_0 = 9.5$ kPa and a resolution of 78 {$\upmu$}m.}
	\label{fig.9500kPa}
\end{figure}

Although exact locations and timings of each detonation re-initiation event differed from simulation to simulation, we found that in most cases detonation re-initiation occurred through a local explosion event that was triggered by the passing of a transverse pressure wave over the interface that separated burned from unburned gases. For example, a detailed sequence of events where detonation re-initiation occurred for $p_0 = 9.5$ kPa is shown in the density gradient evolution of Fig.~\ref{fig.9500kPa}. In Fig.~\ref{fig.9500kPa}a, two triple points have formed due to the propagation of reflected waves from both the top and bottom boundaries of the simulation. These triple points traveled toward each other and eventually collided, as shown in Fig.~\ref{fig.9500kPa}b.  This caused the formation of new reflected waves with increased temperature and pressure behind them (Fig.~\ref{fig.9500kPa}c).  At the same time, a pocket of unburned gas formed behind the various shock dynamics (Fig.~\ref{fig.9500kPa}d). The reflected waves propagated through areas of both shocked and unburned gas as well as the burned gas, and passed through the latter more quickly due to its lower acoustic impedance (Fig.~\ref{fig.9500kPa}e). {The downward reflected shock wave triggered a hot spot near the reaction zone. At the same time, this shock wave, which was traveling in both the burned and unburned gases, triggered a localized explosion on the surface of the interface,} as shown in Fig.~\ref{fig.9500kPa}f.  The subsequent and nearly simultaneous detonation re-initiations along both the Mach and transverse waves are shown in Figs.~\ref{fig.9500kPa}g and h. These transverse detonation waves were self-sustained and continued to propagate to the upper and lower boundaries of the channel until the detonation front was completely re-established. The bulk of the pockets of unburned gas were consumed by the expanding explosion event itself and the newly formed transverse detonations. The remaining gas pockets burned up as deflagrations, which were most likely enhanced by Richtmyer-Meshkov instabilities that arose from the passage of the reflected transverse shock through the pocket surfaces. This sequence of events is in contrast to Bhattacharjee's observations, who speculated that the transverse detonations were formed as a consequence of turbulent burning of the pocket of unburned gas \cite{Bhattacharjee2013b}. Instead, the simulations show that the formation of an explosion event arising from the passing of reflected transverse waves over a burned/unburned gas surface near the Mach shock and after a triple point collision plays a key role in transverse detonation formation.  In the sequence of events observed here, the burn up of the pockets of unburned gas were instead influenced by the passing of the transverse waves after the transverse detonations have formed.

\begin{figure}
	\begin{center}
		\includegraphics[width=88mm]{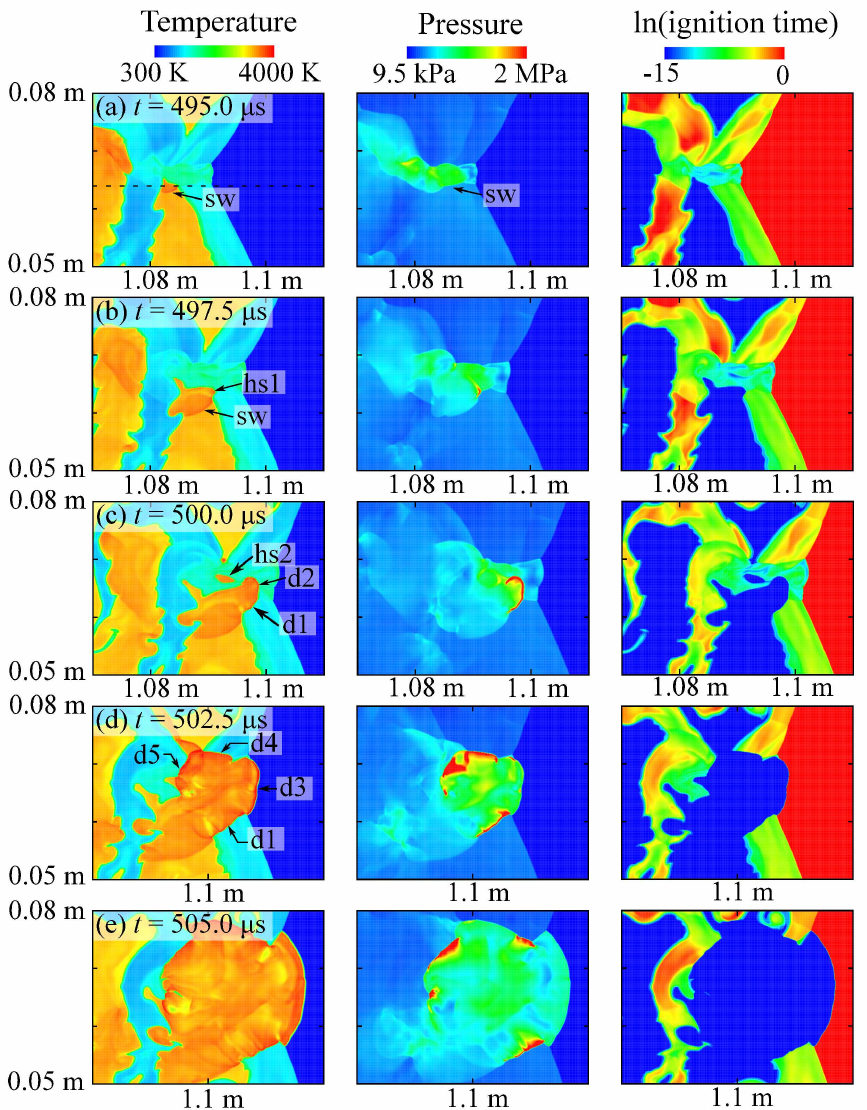}
	\end{center}
	\caption{{Details of temperature, pressure, and ignition delay times for detonation re-initiation of $p_0 = 9.5$ kPa and a resolution of 78 {$\upmu$}m.}}
	\label{fig.9500kPa_detailed}
\end{figure}

\begin{figure}
\centering
\includegraphics[width=88mm]{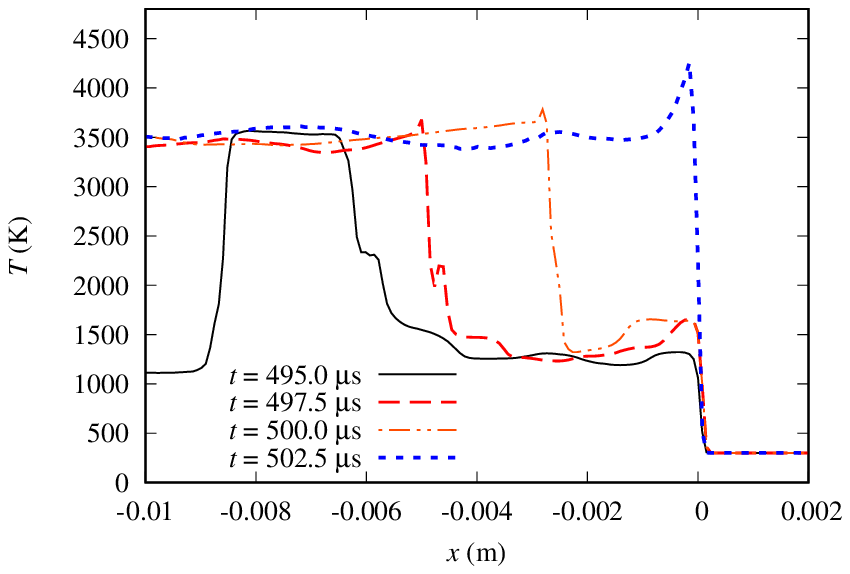}
\includegraphics[width=88mm]{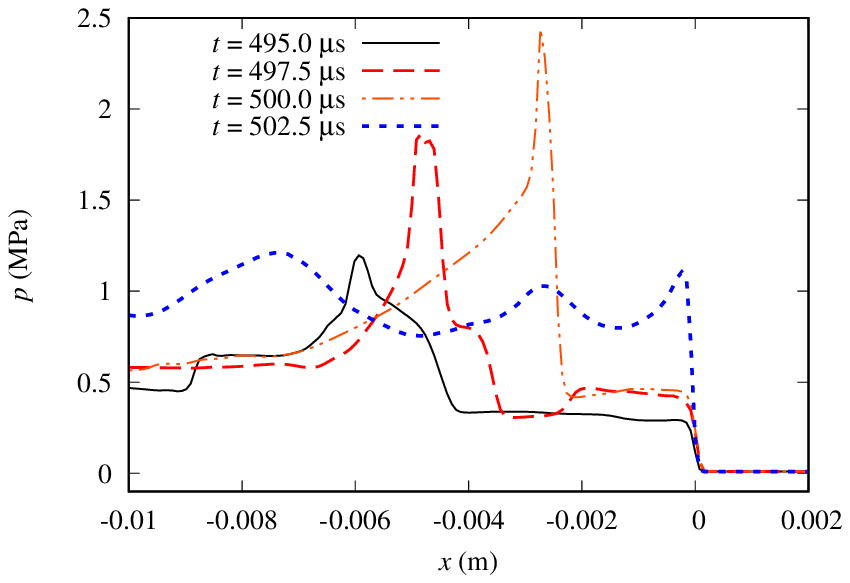}
\includegraphics[width=88mm]{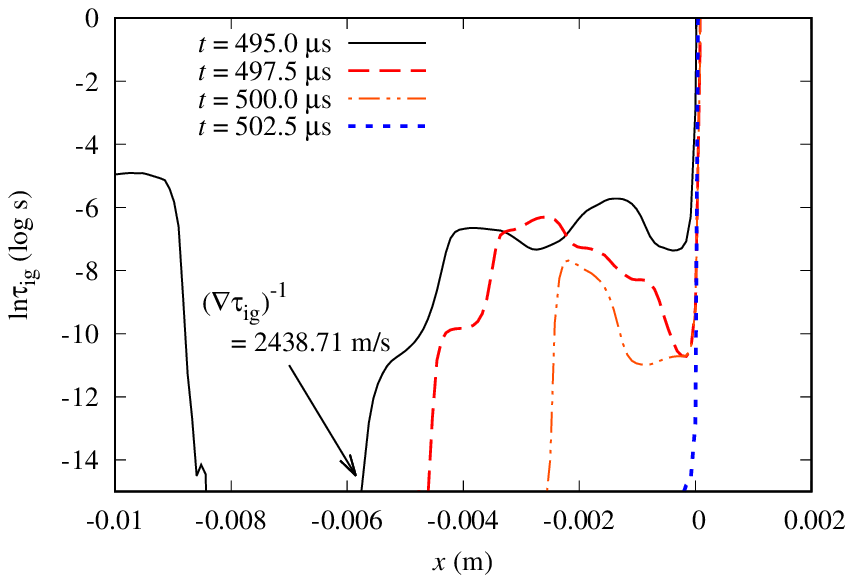}
\caption{{Temperature, pressure, and ignition delay time profiles measured along  along $y=0.064$ m (the dashed line indicated in Fig.~\ref{fig.9500kPa_detailed}a) for $p_0=9.5$ kPa.  Distances are given in the frame of reference of the Mach shock.}}
\label{fig.9500kPa_1Dplots}
\end{figure}

{To gain more clarity on the formation of the detonation waves observed in Fig.~\ref{fig.9500kPa}, detailed temperature, pressure, and local ignition delay time profiles are shown in Fig.~\ref{fig.9500kPa_detailed} for the moments where detonation initiation occurred.  In frame (a), at $t=495$ {$\upmu$}s, the downward propagating transverse shock wave (sw) passed over the burned and unburned gas interface.  This lead to the rapid growth of the existing flame surface, or hot spot (hs1).  At $t=497.5$ {$\upmu$}s, the rapid energy release in this region lead to the localized formation of increased pressure, as seen in the pressure plot of frame (b).  The growth of hot spot (hs1) spread faster in the region of lower ignition delay times, but the increased pressure also directly coupled the rapid chemical reactions to the downward propagating transverse shock wave.  The transverse detonation (d1) thus appears to have been initiated directly by the passing of the transverse shock over the burned/unburned gas interface, which lead to a local pressure amplification and rapid coupling of the shock and reaction zone.  The detonation (d2), on the other hand, appears to have formed through amplification of pressure through the spread of the host pot (hs1) into the gas which contained favorable ignition delay times behind the Mach shock. In fact, the ignition delay times behind the Mach shock was so short, that auto-ignition of new hot spots, such as hot spot (hs2), was possible. Figure \ref{fig.9500kPa_1Dplots} shows the temperature, pressure, and log(ignition delay) profiles at different times along the horizontal dashed line shown in Fig.~\ref{fig.9500kPa_detailed}a, at $y=0.064$ m.  According to the pressure profiles, the explosion hot spot (hs1) clearly experienced a pressure amplification through time.  As the wave propagated against the ignition delay time gradient, shown in the ignition delay time profiles, further amplification of the pressure wave occurred.  The mechanism of detonation initiation thus resembles the well known SWACER (Shock Wave Amplification by Coherent Energy Release) mechanism \cite{LEE1980359}.  In this case, however, the initial explosion was driven and enhanced through Richtmyer-Meshkov instabilities by the passing of an external shock wave over an existing hot spot, and not started by the spontaneous ignition of the gas having minimum ignition delay. Upon measuring the ignition delay time gradient ahead of the reaction wave, we found that the inverse of the ignition delay was $(\nabla \tau_\mathrm{ig})^{-1}\sim2400$ m/s only right before the wave front, at all times. Ahead of the wave, $(\nabla \tau_\mathrm{ig})^{-1}$ was only $\mathcal{O}$(1 to 100) m/s.   This observation is in fact consistent with the past work of Kuznetsov et al.~\cite{Kuznetsov2010} where the $(\nabla \tau_\mathrm{ig})^{-1}$ of the pre-heated mixture during DDT of ethylene--oxygen was also much less than the CJ-detonation speed.  In fact, the recent work of Wang et al.~\cite{WANG2018400} demonstrates that detailed mechanisms are able to permit detonation initiation in much shallower ignition delay gradients compared to simple combustion models (i.e. the one-step combustion model). Although we applied a fairly simple 4-step global combustion model, we have demonstrated its ability to at least mimic a wide spectrum of detailed ignition delay times and steady detonation profiles \cite{Peswani2022}. Although gradients in $\tau_\mathrm{ig}$ were shallow in this case, it is important to point out that non-uniformities did exist in the ignition delay time profiles of Figs.~\ref{fig.9500kPa_detailed} and \ref{fig.9500kPa_1Dplots}, and that such gradients may have promoted the propagation of the reaction wave until a sustained detonation has formed \cite{Oran2007}.  In this case, the detonation front (d2) eventually propagated outward in every direction and reached the Mach shock and upper transverse shock, where the problem became one of detonation transmission from one fluid to another.}

\begin{figure}
	\begin{center}
		\includegraphics[width=88mm]{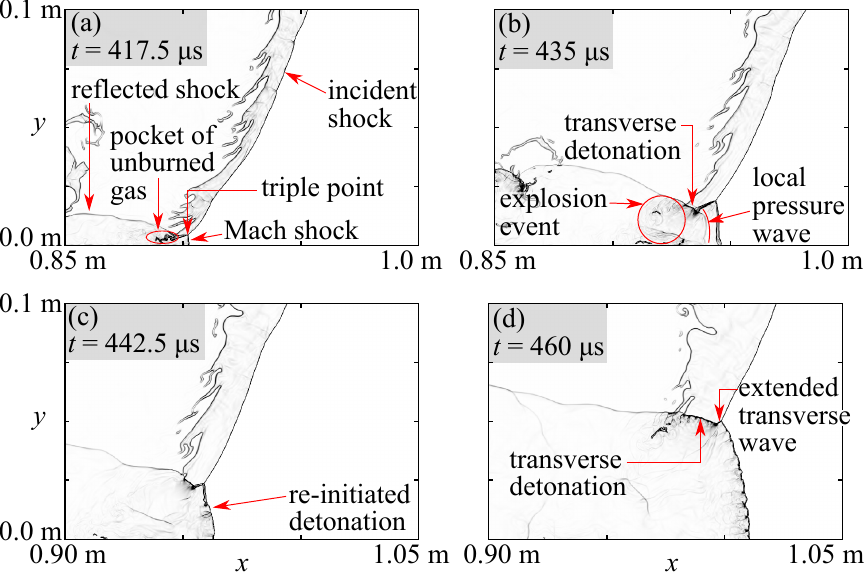}
	\end{center}
	\caption{Density gradient evolution at the moment of detonation re-initiation for an initial pressure of $p_0 = 10.25$ kPa and a resolution of 78 {$\upmu$}m.}
	\label{fig.10250kPa}
\end{figure}

\begin{figure}
	\begin{center}
		\includegraphics[width=88mm]{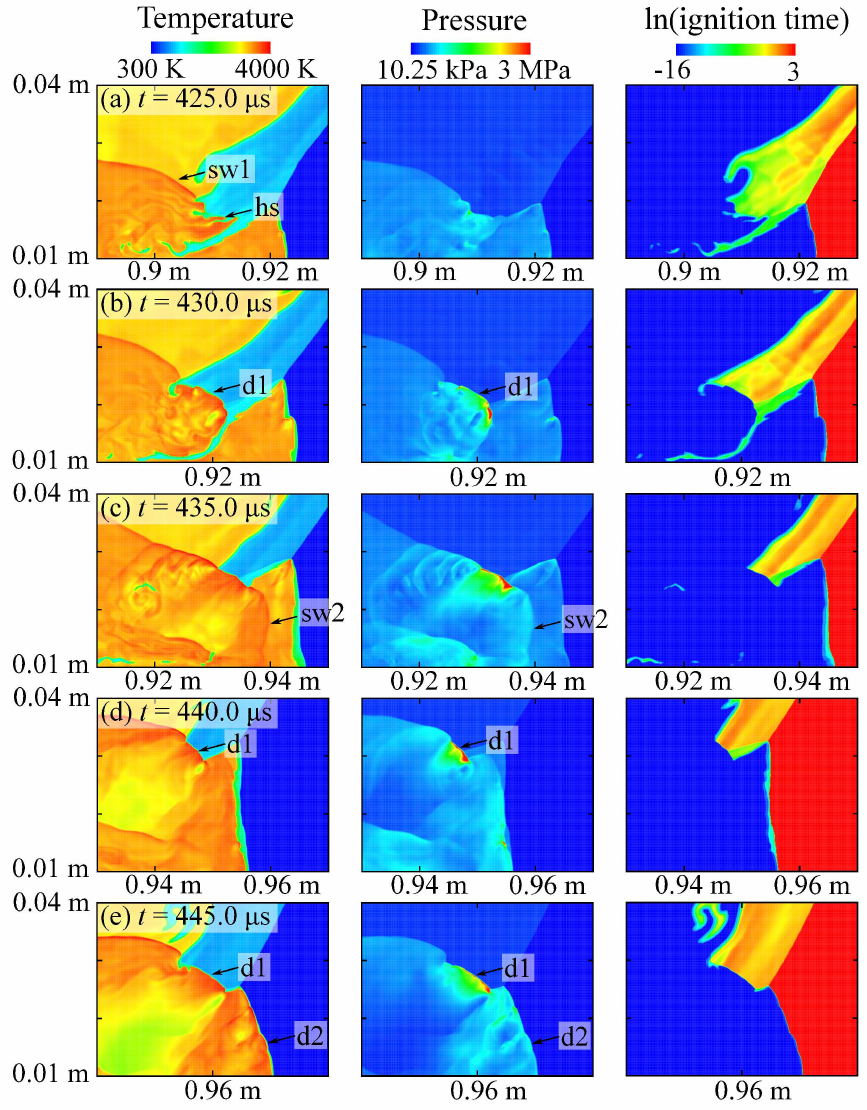}
	\end{center}
	\caption{{Details of temperature, pressure, and ignition delay times for detonation re-initiation of $p_0 = 10.25$ kPa and a resolution of 78 {$\upmu$}m.}}
	\label{fig.10250kPa_detailed}
\end{figure}

The case of $p_0 = 10.25$ kPa had a slightly different sequence of events, as shown in Figs.~\ref{fig.10250kPa} {and \ref{fig.9500kPa_detailed}}. Instead of detonation re-initiation forming behind the collision of two reflected waves propagating toward each other, the detonation re-initiated behind the transverse shock reflection on the bottom boundary. The different outcomes of reflected wave patterns at different pressures is not surprising, as the evolution of shock reflections and triple point locations are likely influenced by the cellular structure of the detonation prior to quenching.  Moreover, it is well known that the detonation cellular structure has a strong dependence on pressure \cite{kaneshige1997detonation}.  Figure \ref{fig.10250kPa}a shows the initial formation of the reflected transverse shock, Mach shock, and triple point. In this case, a flow field that resembled the critical ignition case had formed, where a significant amount of gas had been ignited behind the Mach shock, yet the Mach shock velocity was measured to remain $\sim12\%$ below the CJ velocity prior to detonation re-initiation as shown previously in Fig.~\ref{fig:speed_bottom}.  An explosion event occurred on the surface of the pocket of unburned reactive gas as a result of the passing of the reflected shock wave. This explosion event triggered a transverse detonation wave and also generated local pressure waves that propagated outward toward the Mach shock (Fig.~\ref{fig.10250kPa}b).  The explosion event also directly consumed the pocket of unburned gas. The supported Mach shock transitioned into a self-sustaining detonation wave as shown in Fig.~\ref{fig.10250kPa}c.  At the same time, the transverse detonation wave continued to consume shocked but unburned gas behind the incident shock wave (Fig.~\ref{fig.10250kPa}d). {The details of the detonation initiation, in this case, are revealed in Fig.~\ref{fig.10250kPa_detailed}.  In this case, and much like the 9.5 kPa case discussed previously, the reflected transverse shock (sw1) passed over the burned/unbruned gas interface.  As a result, the reaction rate of hot spot (hs) was enhanced by Richtmyer-Meshkov instabilities, which generated a local pressure rise as observed in the pressure plot of Fig.~\ref{fig.10250kPa_detailed}b.  This lead to the direct rapid coupling of transverse shock and reaction zone, which thus initiated the transverse detonation (d1).  In Fig.~\ref{fig.10250kPa_detailed}c, a shock wave (sw2), generated from the local explosion of the hot spot (hs), propagated towards the Mach shock.  At first this shock travelled in the burned gas, but then initiated the detonation along the Mach shock (d2) through pressure amplification in a short region of shocked and unburned gas that contained gradients in ignition delay times. This can be seen in Figs.~\ref{fig.10250kPa_detailed}d and e.}

\begin{figure}
	\begin{center}
		\includegraphics[width=88mm]{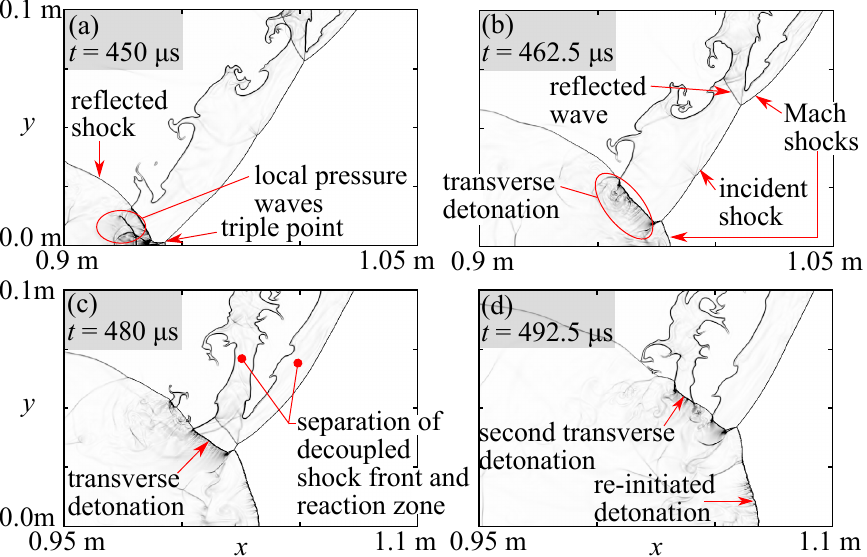}
	\end{center}
	\caption{Density gradient evolution at the moment of detonation re-initiation for an initial pressure of $p_0 = 13$ kPa and a resolution of 78 {$\upmu$}m.}
	\label{fig.13kPa}
\end{figure}

\begin{figure}
	\begin{center}
		\includegraphics[width=88mm]{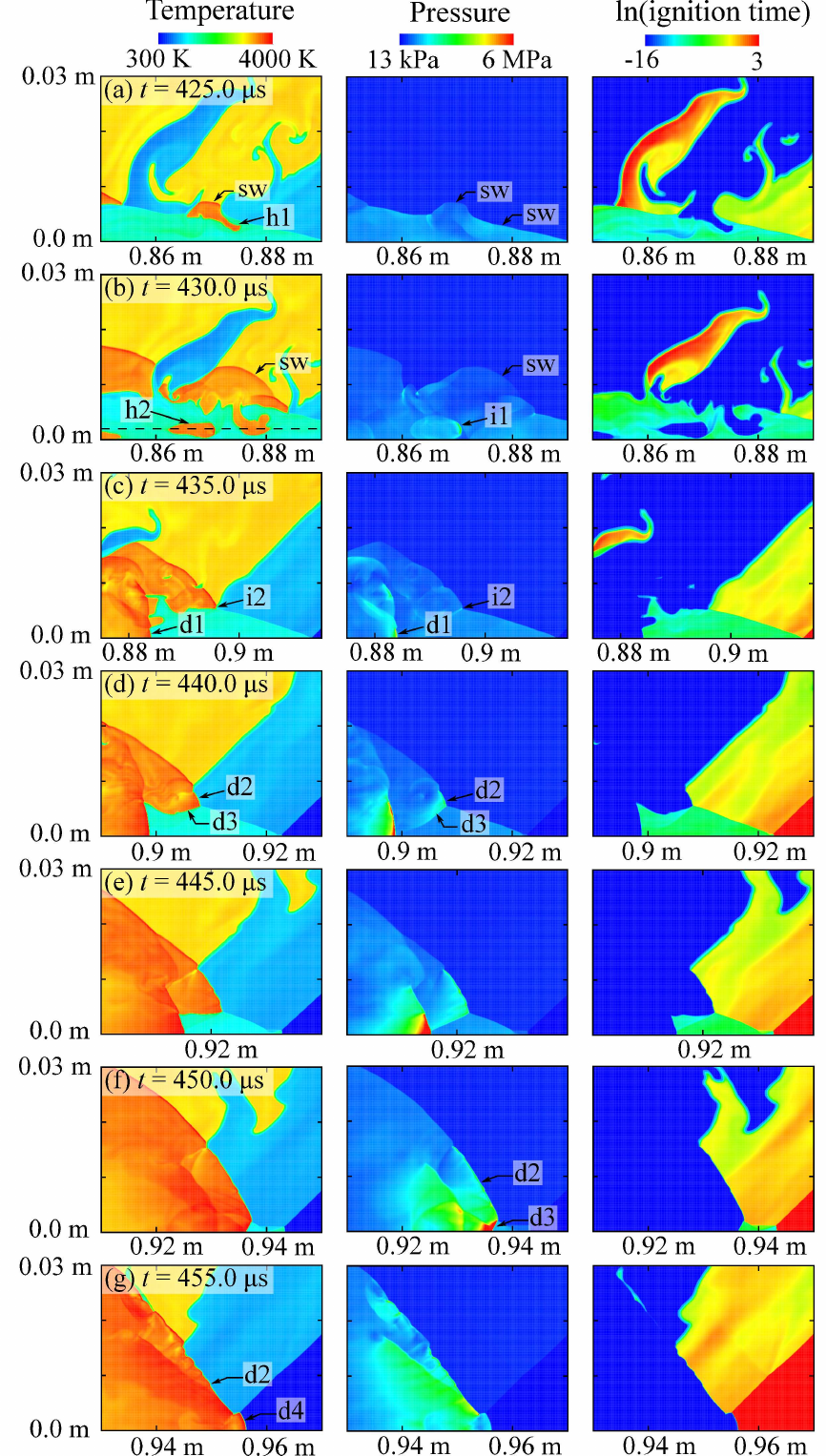}
	\end{center}
	\caption{{Details of temperature, pressure, and ignition delay times for detonation re-initiation of $p_0 = 13$ kPa and a resolution of 78 {$\upmu$}m.}}
	\label{fig.13000kPa_detailed}
\end{figure}

\begin{figure}
\centering
\includegraphics[width=88mm]{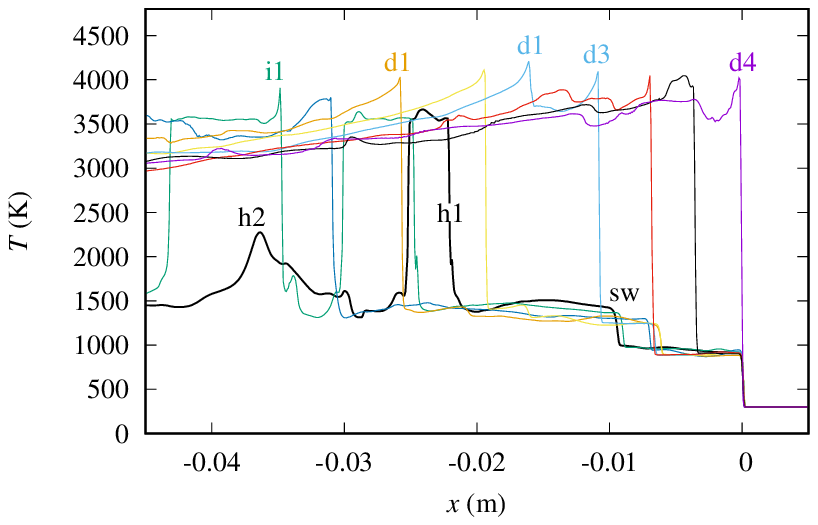}
\includegraphics[width=88mm]{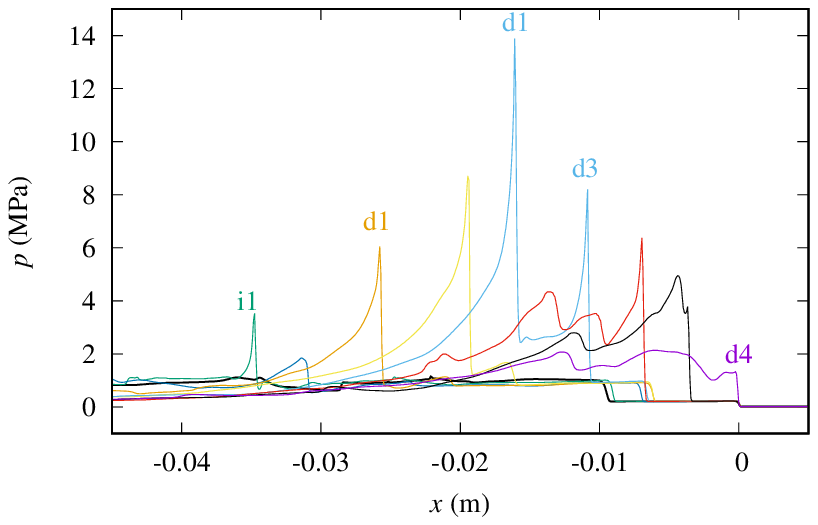}
\includegraphics[width=88mm]{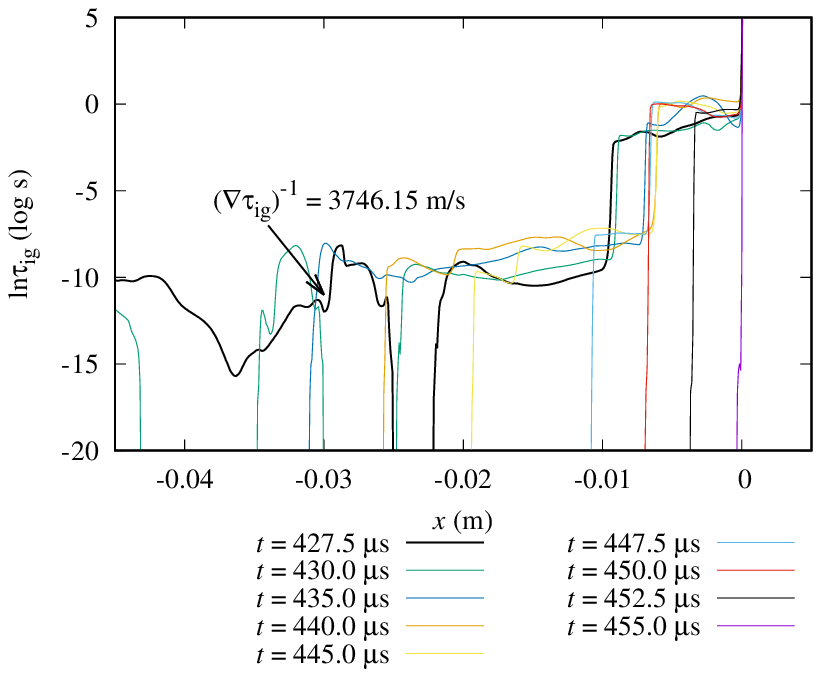}
\caption{{Temperature, pressure, and ignition delay time profiles measured along $y=0.002$ m (the dashed line indicated in Fig.~\ref{fig.13000kPa_detailed}b) for $p_0=13$ kPa.  Distances are given in the frame of reference of the Mach shock.}}
\label{fig.13000kPa_1Dplots}
\end{figure}

At elevated pressures, the transverse detonation was observed to form shortly after from the initial shock reflection on the bottom wall. This is shown for $p_0 = 13$ kPa in the density gradient evolution of Fig.~\ref{fig.13kPa}. This rapid initiation of the transverse detonation wave appears to be similar to the formation previously shown by Lau-Chapdelaine \cite{lau2013numerical}, who observed a transverse detonation initiation directly from the shock reflection using a two-step model at $p_0 = 13.9$ kPa. However, in this past work, the transverse detonation was not self-sustained as it was in this current study.  A main difference in this case compared to the other two cases discussed above is the split in the decoupled shock front and the reaction zone, or a pocket that formed from the shock reflection from the top wall.  This created two zones of shocked yet unreacted gas (Fig.~\ref{fig.13kPa}c). Because of this, two transverse detonations formed that originated from the bottom of the channel  (Fig.~\ref{fig.13kPa}d). The second transverse detonation served to consume the second decoupled reaction zone (the pocket), and disappeared after that.  {The details of the detonation initiation, in this case, are shown in Fig.~\ref{fig.13000kPa_detailed}.  Much like the previous two cases, the reflected transverse shock (sw) passed over a burned/unbruned gas interface, enhancing its combustion and increased reaction rate through Richtmyer-Meshkov instabilities.  This lead to the rapid growth of the shocked hot spot (h1) into the gas with a favourable ignition delay time, as shown in Fig.~\ref{fig.13000kPa_detailed}a.  Much like the 9.5 kPa case, a hot spot (h2) was formed spontaneously in the region of lowest igntion delay, as shown in Fig.~\ref{fig.13000kPa_detailed}b.  In this case, however, the rapid ignition of the newly formed hot spot (h2) was sufficient to trigger a pressure increase locally where shown (i1).  In fact, this hot spot was found to transition to detonation (d1) through the same pressure amplification mechanism previously shown for the 9.5 kPa case.  This is shown in Fig.~\ref{fig.13000kPa_1Dplots}, which shows the temperature, pressure, and log(ignition delay) profiles at different times along the horizontal dashed line shown in Fig.~\ref{fig.13000kPa_detailed}b, at $y=0.002$ m.  In fact, early on at $t=427.5$ {$\upmu$}s, $(\nabla \tau_\mathrm{ig})^{-1}=3746.15$ m/s at the ignition spot (i1).  Since $(\nabla \tau_\mathrm{ig})^{-1}>U_\mathrm{CJ}$, a spontaneous wave was able to form, which eventually developed into the detonation (d1).  In Fig.~\ref{fig.13000kPa_detailed}c another region of increased pressure was generated where the transverse shown (sw) met with an unburned/burned gas interface.  This ignition spot (i2) lead to the direct coupling of shock and reaction zone, which thus initiated detonations (d2) and (d3) shown in Fig.~\ref{fig.13000kPa_detailed}d.  In fact, while (d2) was initiated directly from the rapid compression and energy deposition, (d3) was found to also develop through the pressure amplification mechanism into the gas containing mild gradients of ignition delay times.  Eventually it was detonation (d3) that was able to first reach the Mach shock to initiate the self-sustained detonation (d4).  Although detonation on the Mach shock, in all of the cases presented above, originated by the passing of a transverse shock on a burned/unburned gas interface, we do note that it is possible for a detonation to be initiated by the spontaneous formation of a hot spot through the Zeldovich gradient mechanism \cite{Oran2007,Zeldovich1970}, such as the formation of detonation (d1) discussed here.  In fact, this mechanism of spontaneous wave formation leading to detonation initiation on the Mach shock was found to be the case for $p_0=10.5$ kPa at the 39 {$\upmu$}m resolution (not shown).  In all situations, however, pressure amplification of reactive waves was found to be a common feature in the re-establishment of the detonation wave in the CDR regime.}

In all of these examples, the transverse detonation wave was an important and very common feature in detonation re-initiation. For the onset of detonation, there appears to be a shock compression event that leads to an explosion event. While the sequence in which the transverse detonation forms varies, it is almost always the avenue by which the re-established detonation front extends to the entire domain height, creating a fully established and self-sustained detonation wave.  Finally, we note that transverse detonations also appear to be the main feature through which detonations survive complete quenching in the critical transmission regime, as shown previously in Fig.~\ref{fig.ct}.

\subsection{Detonation re-initiation without the transverse detonation}\label{s.discussNTD}

\begin{figure}
	\begin{center}
		\includegraphics[width=88mm]{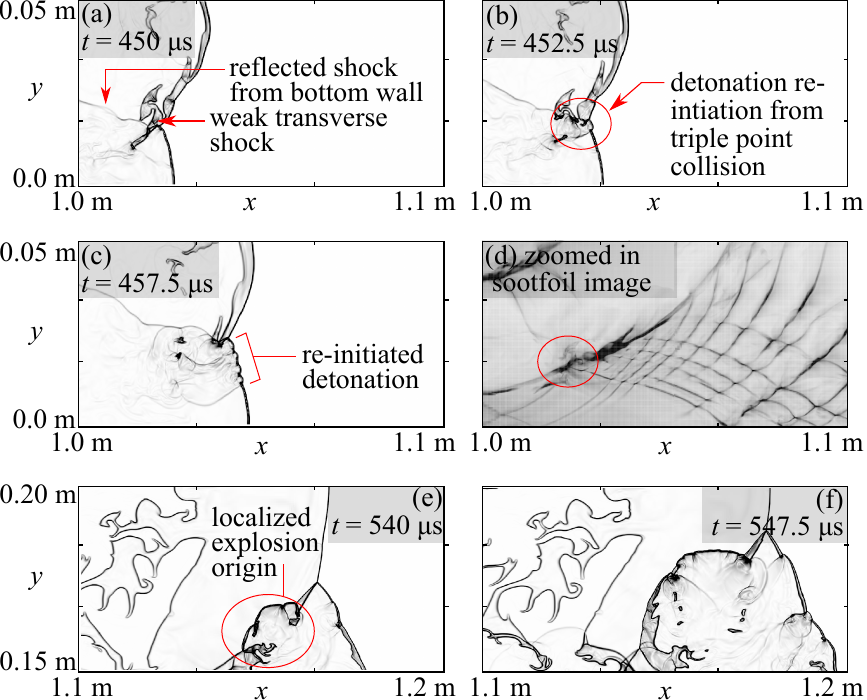}
	\end{center}
	\caption{Density gradient evolution for critical detonation re-initiation without a transverse detonation for $p_0 = 11$ kPa.  {Also shown is a zoomed in portion of the sootfoil (frame d), which corresponds to density gradient frames (a, b, c).}}
	\label{fig.11}
\end{figure}

\begin{figure}
	\begin{center}
		\includegraphics[width=88mm]{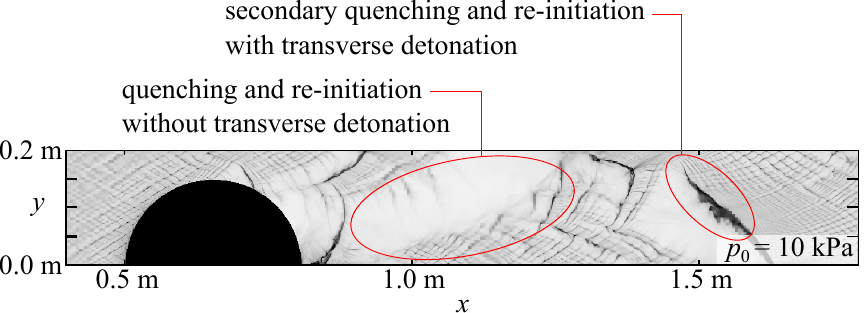}
	\end{center}
	\caption{Numerical soot foil showing CDR with and without transverse detonation for $p_0 = 10$ kPa.}
	\label{fig.soot_10kPa}
\end{figure}

As mentioned earlier, CDR without a transverse detonation was an outcome that was also observed.  Much like the past experiments \cite{Bhattacharjee2013b}, this outcome was not as common as the CDR regime with transverse detonations. Figure~\ref{fig.11} shows the numerical details for $p_0 = 11$ kPa of how the detonation front re-initiated directly {on the Mach shock as the result of a triple point collision of two transverse waves,} however no transverse detonation formed. This could be attributed to the lack of an explosion event {further away from the Mach shock}, which was normally found to be triggered on a decoupled burned/unburned gas surface behind the Mach shock.  We do note, however, that for this case, a localized explosion did eventually occur near the top boundary of the channel around $x=1.15$ m ($S = 1.8$) and $y=0.16$ m, as shown in Fig.~\ref{fig.11}e and f.  This explosion is similar to the early events of CDR with a transverse detonation, however in this case it was not the mechanism that re-initiates the detonation front.  We also note that when CDR without transverse detonation was observed at a lower pressure ($p_0=10$ kPa), such a localized explosion did not occur.  In this case, whose numerical soot foil is shown in Fig.~\ref{fig.soot_10kPa}, global quenching of the wave occurred after re-initiation, which was later re-initiated again but with a transverse wave.  We thus find that CDR  without transverse detonation is a regime that is sensitive to the formation of local explosion events, or quenching of the wave front.

Based on both numerical evidence from this study, and the past experiments \cite{Bhattacharjee2013b}, even though CDR without a transverse wave is possible, a much more likely CDR outcome is that a transverse detonation is triggered, either by a local explosion on a burned/unburned gas surface, or directly by a shock reflection.  This is in contrast to the re-initiation mechanisms observed using past one- and two-step combustion modeling approaches \cite{Bhattacharjee2013,lau2013numerical}.  Thus, it is very likely that in order to capture the event that triggers the transverse detonation, numerically, the ignition response of the combustion model must adapt appropriately to changes in thermodynamic states behind shock waves and reflected shocks. Since the past one- and two-step combustion modeling approaches \cite{Bhattacharjee2013,lau2013numerical} applied calorically perfect gas assumptions, and whose tuning parameters were calibrated only to recover the ignition delay behind a specific predetermined thermodynamic state, errors in temperature, pressure, and also the ignition delay would be expected behind multiple shock dynamics of varying strength.   While Lau-Chapdelaine attributed the lack of a self-sustained transverse detonation to the absence of proper resolution of Richtmyer-Meshkov instabilities \cite{lau2013numerical}, we instead propose this shortcoming to arise due to the lack of ability to accurately model ignition delays in simple combustion models, and therefore such models cannot accurately capture local explosion events behind transverse shock waves at burned/unburned gas surfaces {and their subsequent propagation into gases with non-uniform ignition delay times}.  We believe this is likely why the past numerical attempts have almost always lead to detonation re-initiation along the Mach shock, without the self-sustained transverse detonation wave.

\subsection{Triple point speeds and the transverse detonation strengths}\label{s.tp}

\begin{figure}
\centering
\includegraphics[width=88mm]{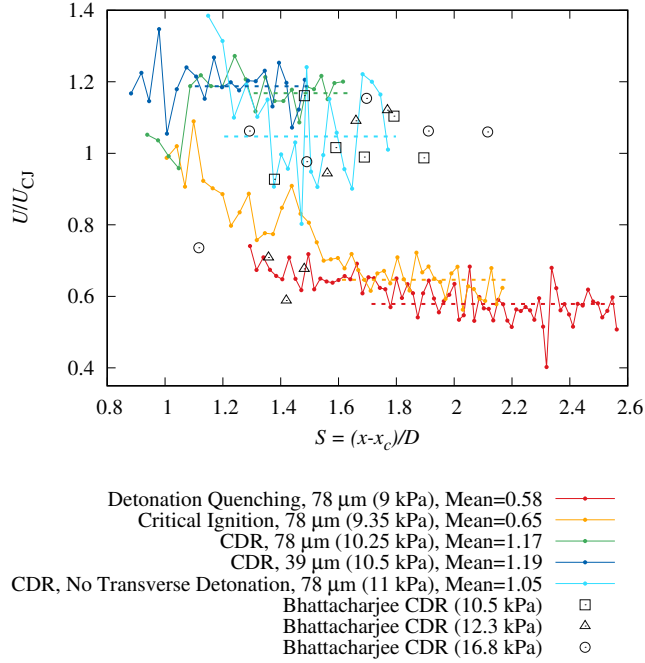}
\caption{Speed of the triple point normalized by the CJ speed vs.~the normalized position along the bottom boundary ($S$) for select cases of Detonation Quenching (DQ), Critical Ignition (CI), Critical Detonation Re-Initiation (CDR), and CDR with No Transverse Detonation (NTD). Also shown are experimental speeds obtained for the Mach shock for three different CDR cases \cite{Bhattacharjee2013b}.}
\label{fig.tp}
\end{figure}

In order to investigate further the quantitative details of the transverse waves, the speed of the triple point was measured for a few different cases. Figure \ref{fig.tp} shows the magnitude of the triple point velocity normalized to the CJ speed ($U_\mathrm{tp}/U_\mathrm{CJ}$) vs.~its normalized distance along the channel from the cylinder center ($S$).  For detonation quenching ($p_0 = 9$ kPa), the triple point clearly travels much below CJ speed from its formation to a distance in $S$ which corresponds to the triple point reaching the top boundary, with a mean normalized speed of $U_\mathrm{tp}/U_\mathrm{CJ} = 0.58$. For the critical ignition case at $p_0=9.35$ kPa, the triple point first travels near CJ speed, driven by intense combustion behind the newly formed Mach shock.  Eventually the speed of the triple point, and thus the absolute speeds of the transverse and Mach waves, settle to normalized speeds closer to that of the detonation quenching case by the time the triple point reaches $S\sim1.6$, with a mean speed beyond this distance of $U_\mathrm{tp}/U_\mathrm{CJ} = 0.65$. Included in the graph are two cases of critical detonation re-initiation ($p_0 = 10.25$ and $10.5$ kPa), from two different resolutions (78 and 39 {$\upmu$}m respectively). Both cases have very similar behavior, with the triple point speed remaining overdriven through the entire re-establishment of the detonation wave, with mean values of $U_\mathrm{tp}/U_\mathrm{CJ} = 1.17$ and $U_\mathrm{tp}/U_\mathrm{CJ} = 1.19$ for 78 {$\upmu$}m and 39 {$\upmu$}m respectively, with less than a 2\% difference from each other.  Finally, CDR with no transverse detonation shows a triple point speed that oscillates in magnitude above and below the CJ-velocity through the entire domain height, with a mean speed of $U_\mathrm{tp}/U_\mathrm{CJ} = 1.05$.  As a reference, experimental measurements of the Mach shock speeds for CDR at three different pressures from Bhattacharjee \cite{Bhattacharjee2013b} have been included, whose estimated speeds were found to be in the range of 1.05$U_\mathrm{CJ}$ to 1.17$U_\mathrm{CJ}$.  The degree of overdrive experienced by the wave front when a transverse detonation wave is present is consistent between the simulations and the past experiments.  Since the triple point speed of CDR without a transverse detonation travels close to the CJ-speed, while CDR is always overdriven, we can attribute the overdriven state of the Mach shock and triple point to the presence of the transverse detonation.  It is very likely that the rapid energy release and subsequent expansion of the products behind the transverse detonation act as a piston to overdrive the Mach stem through the re-initiation process.  This mechanism also acts to explain the overdriven speed measurements and pressure amplification recently observed in experiments of Chin et al.~\cite{Chin2020} for detonation quenching and re-initiation following a critical geometric area expansion.

\begin{figure}
	\begin{center}
		\includegraphics[width=88mm]{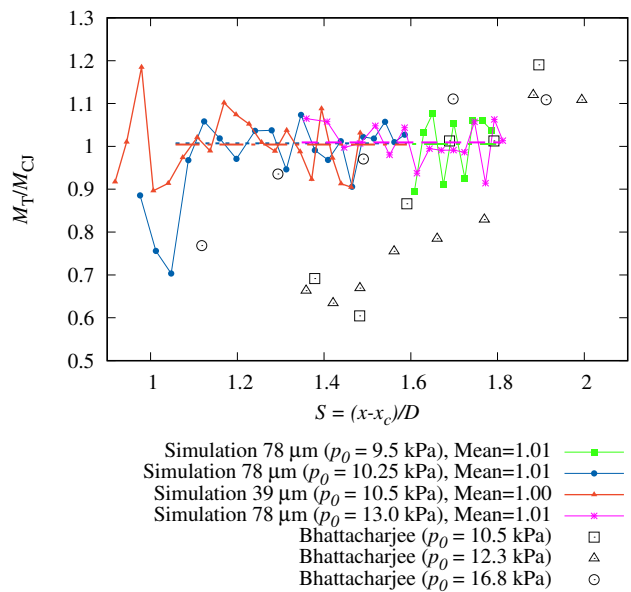}
	\end{center}
	\caption{Strength of the transverse detonation normalized by the CJ Mach number in the shocked and unburned gas ($M_\mathrm{T}/M_\mathrm{CJ}$) vs.~the normalized position along the bottom boundary ($S$) for select cases of  Critical Detonation Re-Initiation (CDR), and compared to experimental estimates \cite{Bhattacharjee2013b}.}
	\label{fig.transverseStrength}
\end{figure}

In order to estimate the strength of the transverse detonation, we considered the velocity vector of the triple point relative to the velocity vector of the shocked gas in front of the transverse detonation.  Here, we chose to use the triple point velocity as a reference point to measure, as its absolute lab-frame velocity can be assumed to be close to that of the transverse detonation itself.  This also removed uncertainty in finding a suitable reference point attached to the transverse wave itself.  To determine the velocity of the shocked gas in front of the transverse wave, we considered a sample space of approximately 12.75 mm in the $x$-direction by 7.875 mm in the $y$-direction, consisting of, on average, 16,000 points on the finest refinement level.  From the data points in this range, mass weighted averages were obtained for the shocked gas velocity components, denoted as $\tilde{u}_s$ and $\tilde{v}_s$.  Here the mass weighted average of a scalar was computed from
\begin{equation}
    \tilde{\phi}=\frac{\overline{(\rho\phi)}}{\overline{\rho}},
\end{equation}
where $\phi$, represents the scalar of interest that was averaged, and the over-line represents an ensemble average.  Once the mass-weighted velocity of the transverse wave was determined relative to the shocked gas, its strength was obtained by normalizing the velocity to the mass-weighted average speed of sound in the shocked gas, i.e.
\begin{equation}
M_T=\sqrt{(u_\mathrm{tp}-\tilde{u}_s)^2+(v_\mathrm{tp}-\tilde{v}_s)^2}/{\tilde{c}_s}.
\end{equation}
Then from the ensemble averaged density and pressure, $\overline{\rho}$ and $\overline{p}$, the Mach number of the CJ-solution ($M_\mathrm{CJ}$) associated with the shocked and unburned state was determined.  We thus present ($M_T/M_\mathrm{CJ}$) vs. normalized distance from the center of the cylinder ($S$) for several different initial pressures in Fig.~\ref{fig.transverseStrength}.  In all CDR cases simulated, including those not shown, we found that the transverse detonation was in fact a CJ-detonation (within 1\%).  Also shown for comparison are estimated transverse detonation strengths from Bhattacharjee's experiments \cite{Bhattacharjee2013b} for three different pressures.   Although Bhattacharjee had estimated the transverse detonation strength to vary from 0.6 to 1.2$M_\mathrm{CJ}$, we note that significant errors likely existed in the experimental estimation of the sound speed from schlieren images.  Moreover, Bhattacharjee's estimate did not consider the relative difference in the velocity vectors of the triple point and the gas behind the incident shock.

\begin{figure}
	\begin{center}
		\includegraphics[width=88mm]{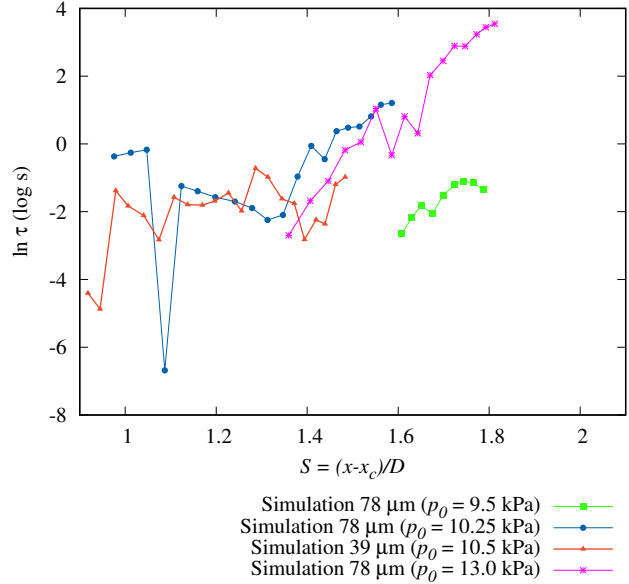}
	\end{center}
	\caption{{Ignition delay time of the shocked and unreacted gas in front of the transverse detonation vs.~the normalized position of the triple point along the bottom boundary ($S$) for select cases of  Critical Detonation Re-Initiation (CDR).}}
	\label{fig.igntime_vs_S}
\end{figure}

{Finally, the ignition delay times were calculated for the same sample windows used to estimate the transverse detonation strengths.   Figure~\ref{fig.igntime_vs_S} thus shows the calculated ignition delay of the shocked and unreacted gas in front of the transverse detonation vs.~the normalized position of the triple point along the bottom boundary ($S$) for several CDR cases.  Although the transverse detonations were found to be CJ-detonations, in the shocked and unreacted gas, we do note the presence of an ignition delay time gradient.  In all cases there is an increasing trend in the ignition delay time as the detonation wave evolves.  It is possible that this ignition delay time gradient helps to sustain the transverse detonation wave through the global detonation re-initiation process, and should be investigated in more detail to confirm if such an ignition delay time gradient is in fact required to sustain the transverse detonation.  We do note here, however, that such a gradient in the ignition delay time exists since the incident shock of the quenched detonation weakens with time, causing lengthened induction lengths of the unreacted gas as more time passes.}

\section{Conclusion}

In this investigation we applied a thermochemically derived four-step global combustion model \cite{Zhu_2012} to investigate critical detonation attenuation and the role of transverse detonations during its re-establishment following its interaction with obstacles \cite{Bhattacharjee2013}.  Our simulations have demonstrated that application of this minimal global combustion model is able to capture the sustained transverse detonation feature in this scenario, unlike past applications of simple one- and two-step combustion schemes \cite{Bhattacharjee2013,lau2013numerical}.  We attribute this to the fact that the relatively simple four-step model used contains an adequate description to permit the correct ignition {and thermodynamic state} response when changes in temperature and pressure occur \cite{Peswani_IMECE2020,Peswani_ussci2021,Peswani2022}, i.e.~behind shocks and reflected shocks.  This appears to be required not only to capture the transverse detonation, but also to capture the less frequent situations where detonation re-initiation occurs without a transverse detonation.  {In both of these cases, accurate treatment of ignition delay time behind shock compression is important.  Perhaps the past applications of one- and two-step combustion models to this scenario did not contain sufficiently steep gradients in ignition delays times to trigger or sustain transverse detonations.  Since the 4-step model applied gives rise to ignition delay times that effectively respond appropriately to changes in the thermodynamic state, when compared to delailed chemistry, detonations can likely form in shallower ignition delay time gradients compared to the past one- and two- step modeling approaches. In addition to this, we}  also acknowledge that closure of turbulent mixing is also equally important for capturing critical ignition associated with the lower pressure limits of the critical regime.  {In future work, we recommend the coupling of the four-step combustion model to the compressible linear eddy model for large eddy simulation approach \cite{maxwell_2017}.  For now, however, we draw our conclusions} from Euler simulations where turbulent mixing is implicitly controlled through resolution of the numerical scheme.

{In this work, we have found that there exists a relationship between the outcome and the cell size and geometry involved.  In this investigation, we found that a range of critical outcomes was possible when $3 \le (d_\mathrm{H}/\lambda)_\mathrm{crit} \le 6$, where $(d_\mathrm{H}/\lambda)$ is the gap size to mixture cell size ratio.  In future work, influences on gap size and cylinder radius should be explored in a parametric investigation.  For the critical detonation re-initation outcome, we have clarified that one principal mechanism through which transverse detonations and detonations along the Mach shock can form is through pressure amplification of reaction zones at burned and unburned gas interfaces behind Mach shocks, and in the presence of ignition delay time gradients.  In this mechanism, the passing of the transverse shock wave over the burned and unburned gas interface leads to enhanced combustion rates through Richmyer-Meshkov instabilities, which generates the pressure necessary to amplify into a coupled shock and reaction zone, or detonation.   These detonations are also possible to form through spontaneous ignition of the gas, i.e.~from a hot spot formed by the passing of transverse shocks in regions of lowest ignition delay times, which can ultimately form through the Zeldovich gradient mechanism \cite{Zeldovich1970}.  When transverse detonations do not form, it is possible for detonation re-initiation to occur on a Mach shock directly through a triple point collision.}  However, this outcome is not as common as the former, and was found to be sensitive to local explosions, or quenching.  {Also, since higher pressures have more triple points that can survive the expansion from the obstacle, and since direct initiation of a detonation on the Mach shock following a triple point collision does not produce transverse detonations, this likely explains why transverse detonations are only observed at critical low pressures. In addition to all of this,} it was confirmed that transverse detonations are indeed CJ-detonations, and whose presence allows for the detonation along the Mach stem to be overdriven.  Finally, our simulations have revealed that while pockets of unburned gas may exist when transverse detonations occur, it is not the direct burn-up of these pockets that give rise to transverse detonations as previously suspected.  Instead, the pockets of unburned gas are consumed by their own deflagrative burning, or by the passing of such transverse detonation waves.

\section*{Acknowledgements}

This research was enabled in part by high performance computing resources provided the Core Facility for Advanced Research Computing at Case Western Reserve University (CWRU).  GF was supported for this work by a research scholarship provided by the CWRU SOURCE SURES program with additional funding provided by the Case Alumni Association, and also through an Ohio Space Grant Consortium Undergraduate STEM Scholarship and a Selected Professions Fellowship from the American Association of University Women.

\appendix
\section{Effect of resolution and internal boundary conditions on the inert gasdynamic evolution}
\label{sec.InternalGeometry}

{In this investigation, to handle the presence of the internal cylinder geometry, the straight-forward \emph{staircase} type of boundary was constructed within the computational domain.  Specifically, cells are marked as either being a fluid or a solid depending on their location.  While simple to implement, this method is known to introduce artificial roughness to the flow and also introduces nonphysical waves which originate from the surface \cite{Cangellaris1991}.  As a result, local errors of $\mathcal{O}(1)$ may also appear near the surface \cite{Haggblad2014}.  A common alternative in a Cartesian grid-based framework would have been to adopt an embedded boundary technique \cite{Xu1997}.  However, this method is not necessarily conservative, and can result in different Mach, transverse, and incident shock configurations when compared to shock-wedge simulations where boundaries are aligned with the grid itself \cite{lauChapdelaine2013shockwaves,lauChapdelaine2019}.  The cut-cell approach \cite{Tucker2000} is another popular method for Cartesian grids, but the possibility may arise for modified cells near the boundary to become too small, which may lead to numerical instability.  In the end, we chose to use conventional staircase boundaries for two main reasons:  (1) Numerical stability of the scheme was ensured, and (2) conservativity was satisfied such that undesirable flow leakage was avoided.  Also, since errors originating from staircase boundaries are local in nature, it has been suggested that such errors may be neglected in applications where flow fields along the boundary are not the main focus \cite{Haggblad2014}.  In this section, we examine the influence of the discrete internal boundary conditions on the evolved unsteady flow fields of inviscid and inert shock-cylinder interactions at different resolutions.  Here, we considered the same domain previously shown in Fig.~\ref{domain}, except that the solution for a shock travelling at 2317.84 m/s was imposed at $x=0.4$ m, which corresponded to the CJ-shock speed solution with $p_0=10$ kPa, and the left boundary prescribed was instead a zero-gradient boundary condition.}

\begin{figure}
	\centering
	\includegraphics[width=88mm]{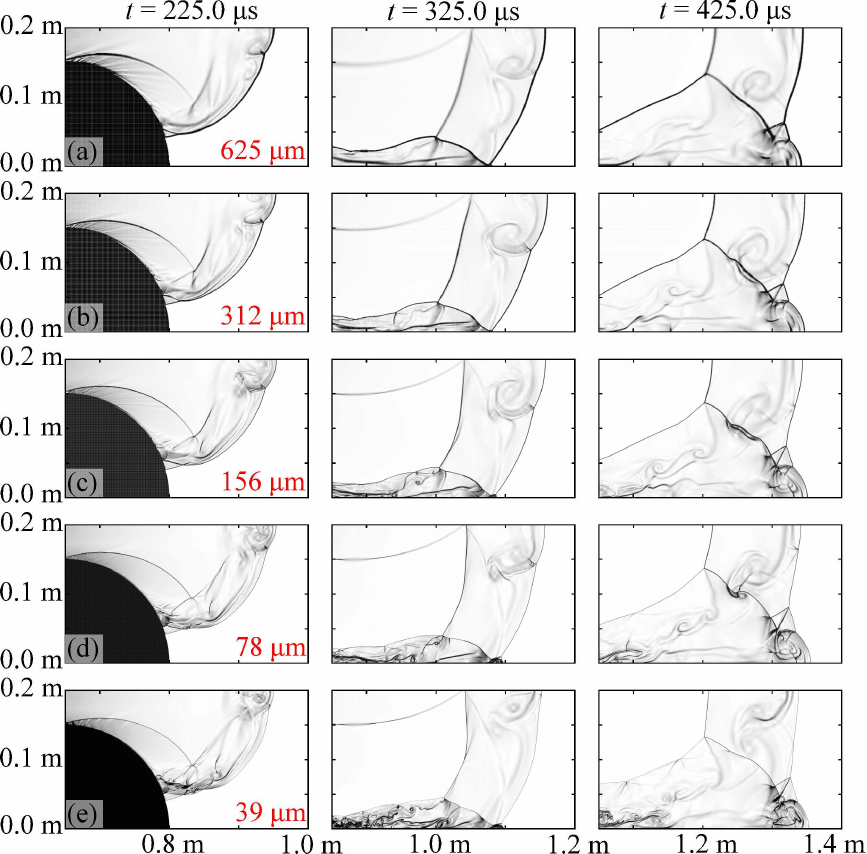}
	\caption{{Density gradient fields of inert simulations at different resolutions, and simulations times.}}
	\label{fig:Inert_shockEvolution}
\end{figure}

\begin{figure}
	\begin{center}
		\includegraphics[width=88mm]{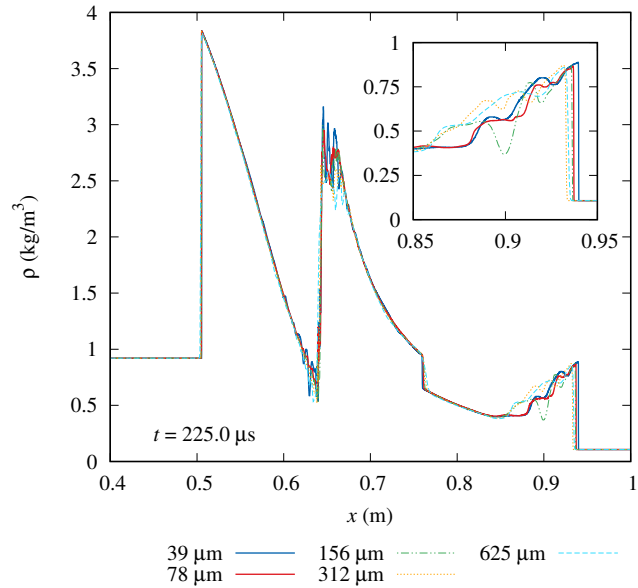}
	\end{center}
	\caption{Density profiles obtained along $y=0.15$ m at $t=225$ {$\upmu$}s for all resolutions.}
	\label{fig.Inert_densityProfiles}
\end{figure}

{Figure \ref{fig:Inert_shockEvolution} shows the instantaneous density gradient fields obtained at three different simulation times ($t=225$ {$\upmu$}s, $t=325$ {$\upmu$}s, and $t=425$ {$\upmu$}s) for all grid resolutions considered in this study:  39 {$\upmu$}m, 78 {$\upmu$}m, 156 {$\upmu$}m, 312 {$\upmu$}m, and 625 {$\upmu$}m.  As can be seen at early times, at $t=225$ {$\upmu$}s, spurious waves were indeed observed near the internal staircase boundary, at all resolutions.  This was more evident in Fig.~\ref{fig.Inert_densityProfiles}, which displays the density profiles obtained along $y=0.15$ m at $t=225$ {$\upmu$}s for all resolutions.  Errors in the density field were found to be the most significant, to leading order, near the top of the obstacle (i.e.~$0.6<x<0.7$ m).  However, these errors do not appear to have heavily influenced the downstream density profiles or shock locations, except maybe by a few mm as shown in the zoomed in portion of Fig.~\ref{fig.Inert_densityProfiles}.  It was likely that errors originating from the internal boundary were damped downstream from the obstacle through the artificial viscosity associated with the HLLC solver used \cite{Toro2009}.  At later times, $t=325$ {$\upmu$}s and $t=425$ {$\upmu$}s in Fig.~\ref{fig:Inert_shockEvolution}, all of the various gasdynamic features that evolved downstream from the obstacle were preserved at all resolutions.  This included the formation of a triple point, bifurcated Mach shock, transverse shock structures, and backward facing shocks.  In the reactive case, it is possible for errors originating from the boundary to influence the local flame development behind quenched zones in critical case, especially near the obstacle.  However, the Euler scheme adopted does not explicitly account for diffusion terms, and is therefore sensitive to changes in resolution anyways.  Based on these observations, we believe that it is unlikely that the internal boundary conditions applied would have significantly influenced the spectrum of outcomes observed beyond the typical resolution related errors associated with the Euler scheme applied.}

\section*{References}

\bibliography{Floring2022}

\end{document}